\documentclass[manuscript]{acmart}
\AtBeginDocument{%
  }

\usepackage[ruled,vlined,linesnumbered]{algorithm2e}

\let\oldnl\nl
\newcommand{\nonl}{\renewcommand{\nl}{\let\nl\oldnl}}

\usepackage{pifont}
\usepackage{bm}
\usepackage{subfigure}
\usepackage{array}
\usepackage{tabularx}
\usepackage{booktabs}
\usepackage{makecell}
\begin{document}

\title[Indexing and Data Distribution in Hybrid DDR--CXL Databases]{Indexing Designs and Adaptive Data Distribution Optimization for In-Memory Databases on Hybrid DDR--CXL Memory}

\author{Huijie Cao}
\affiliation{%
  \institution{East China Normal University}
  \city{Shanghai}
  \country{China}
}
\email{hjcao@stu.ecnu.edu.cn}
\orcid{0009-0004-9777-2513}

\author{Shun Yang}
\affiliation{
  \institution{East China Normal University}
  \city{Shanghai}
  \country{China}
}
\email{syang@stu.ecnu.edu.cn}
\orcid{0009-0007-8157-0396}

\author{Yinan Zhang}
\affiliation{
  \institution{East China Normal University}
  \city{Shanghai}
  \country{China}
}
\email{ynzhang@stu.ecnu.edu.cn}
\orcid{0009-0009-2398-9185}

\author{Huiqi Hu}
\affiliation{
  \institution{East China Normal University}
  \city{Shanghai}
  \country{China}
}
\email{hqhu@dase.ecnu.edu.cn}
\orcid{0000-0001-5220-3166}

\author{Xuan Zhou}
\affiliation{
  \institution{East China Normal University}
  \city{Shanghai}
  \country{China}
}
\email{xz@dase.ecnu.edu.cn}
\orcid{0000-0001-8426-9450}

\author{Weining Qian}
\affiliation{
  \institution{East China Normal University}
  \city{Shanghai}
  \country{China}
}
\email{wnqian@dase.ecnu.edu.cn}
\orcid{0000-0002-4132-8630}

\renewcommand{\shortauthors}{H. Cao et al.}

\begin{abstract}
CXL memory offers a new way for in-memory databases to expand the available memory capacity of a single node, but has higher access latency and lower bandwidth than local DDR memory. To exploit both the capacity advantage of CXL memory and the performance advantage of DDR memory, databases must jointly consider indexing design and data distribution. Different indexing designs lead to different data access and migration paths and constrain how indexes and tuples are distributed across the two memory tiers. These differences result in different trade-offs in runtime performance, DDR memory efficiency, and system integration complexity, making the actual impact of each indexing design on database performance difficult to determine directly. Indexes and tuples may also have different access characteristics. Placing them in the same memory tier may reduce DDR memory efficiency, whereas managing them separately increases access tracking and distribution decision overheads.

To address these issues, we propose five indexing designs for hybrid DDR--CXL databases based on two approaches: treating the two memory tiers as a unified data space or managing them separately. We systematically analyze the differences and trade-offs among these designs and their underlying causes. We further design an adaptive data distribution optimization method for single-index designs in which the database explicitly manages data distribution. The method manages index objects and tuples separately and uses hierarchical hotness tracking and an index object residency policy to reduce metadata management overhead. It extends benefit-aware admission and randomized eviction to determine and adjust the distribution of the two object types according to their respective expected benefits and memory footprints. We implemented these designs in a hybrid DDR--CXL database prototype and evaluated them using YCSB. Experimental results show that the single-index design with direct addressing achieves the best performance under most workloads and system configurations. The dual-index design provides performance advantages mainly when DDR memory capacity is limited or accesses are highly concentrated, and has relatively low system integration complexity. The proposed adaptive data distribution optimization method improves DDR memory efficiency and system performance, enabling single-index designs to achieve up to $1.59\times$ the throughput of their corresponding implementations with non-separated management.

\end{abstract}



\maketitle

\section{Introduction}
In-memory databases store data in main memory to avoid storage I/O overhead and provide low latency and high throughput \cite{im1,im2,im3}. However, as the amount of data managed by applications continues to grow, local memory capacity becomes increasingly insufficient to accommodate the entire database. Expanding local DDR memory is constrained by the number of DIMM slots and memory channels \cite{ddr2,ddr1}, while high-density memory modules remain costly. Compute Express Link (CXL) provides an alternative for expanding memory capacity within a single node. Memory connected through CXL.mem can be incorporated into the host address space and accessed by the CPU using ordinary load and store instructions \cite{cxl}. This enables databases to use local DDR memory and CXL memory together to expand their available memory capacity.

Nevertheless, CXL memory has higher access latency and lower bandwidth than local DDR memory \cite{ddr2,ev1,ev2}. Request execution typically involves an index lookup followed by tuple access. Different indexing designs determine which indexes are used for lookup, which memory tiers are accessed, and how tuple addresses are resolved, resulting in different data access paths \cite{Hao,two}. During data distribution adjustment, the indexing design also affects which objects must be migrated and which references must be maintained, resulting in different data migration paths \cite{Tiered-Indexing,SIDLE}. These differences affect request execution overhead and data migration overhead, respectively, and hence runtime performance. The indexing design further constrains whether an index and the tuples it manages must reside in the same memory tier and which objects can be stored in DDR memory. These constraints, however, do not determine the actual distribution of indexes and tuples. The database must still determine their distribution according to their access characteristics and adjust it as those characteristics change \cite{hotspot_11,hybridtier}. Maintaining high performance while expanding memory capacity therefore requires considering both indexing design and data distribution.

First, an indexing design affects multiple processes, including request execution and data distribution adjustment. Different designs involve complex trade-offs among the runtime overheads of these processes: reducing one overhead may increase others. Overall performance depends on the combined effect of these overheads, making the actual performance impact of an indexing design difficult to determine directly. Moreover, data access characteristics, memory capacity, and memory access performance change the extent to which each overhead affects overall performance. As a result, performance differences between indexing designs vary with these conditions. Understanding these overheads and their combined impact under different workload and memory conditions is therefore essential to indexing design.

Second, the performance of an indexing design also depends on how indexes and tuples are distributed across the two memory tiers. Indexes and tuples may have different access characteristics, so placing them together in the same memory tier may reduce DDR memory efficiency. Separate management, however, increases access tracking and distribution decision overheads. It also requires distribution decisions to account for potential differences in memory footprint, access cost, and cross-tier migration cost between indexes and tuples. These differences make it difficult to determine their priority for DDR memory allocation based on access frequency alone.

To address these challenges, we propose five indexing designs and an adaptive data distribution optimization method for hybrid DDR–CXL in-memory databases. The five designs either treat DDR and CXL memory as a unified data space or manage the two memory tiers separately. A systematic analysis characterizes their differences and trade-offs in runtime performance, DDR memory efficiency, and system integration complexity. It further examines the underlying causes.

We further design an adaptive data distribution optimization method for single-index designs in which the database explicitly manages data distribution. The method manages index objects separately from tuples to adapt to their different access characteristics. To control the additional metadata management overhead, it employs hierarchical hotness tracking and an index object residency policy to limit the set of objects requiring continuous tracking and distribution adjustments. It also extends the benefit-aware admission and randomized eviction mechanisms \cite{nephele} to determine and adjust the distribution of index objects and tuples according to their respective expected benefits and memory footprints.

We implemented the indexing designs and the distribution optimization method in a hybrid DDR–CXL database prototype and evaluated them using YCSB \cite{ycsb}. With the proposed distribution optimization, the single-index design with direct addressing achieves the highest throughput in most evaluated workloads and system configurations. The dual-index design provides higher throughput in some configurations with limited DDR capacity or high access skew, and has low system integration complexity. For the single-index designs with explicit distribution management, the optimization improves DDR memory efficiency and achieves up to 1.59× the throughput of the corresponding implementations using non-separated management, which keeps each index object and its associated tuple in the same memory tier.

The main contributions are as follows:
\begin{itemize}
    \item Five indexing designs for hybrid DDR–CXL memory databases, together with a systematic analysis of their differences, trade-offs, and underlying causes. (Section 3)
    \item An adaptive data distribution optimization method that manages index objects and tuples separately while controlling metadata management overhead. (Section 4)
    \item A comprehensive experimental evaluation of the indexing designs and distribution optimization under different workloads, memory capacities, and CXL memory access latencies. (Section 5)
\end{itemize}

Section 2 discusses related work. Section 3 presents the indexing designs and their analysis, followed by adaptive data distribution optimization in Section 4. Section 5 reports the experimental evaluation, and Section 6 concludes the paper.

\section{Related Work}
\label{sec:related-work}

Research on indexing in tiered storage spans memory--disk systems \cite{LeanStore,anti,Siberia,two,Tiered-Indexing,Bf-Tree,Hao}, DRAM--persistent memory systems \cite{FOEDUS,FPTree,HiKV,DPTree,BonsaiKV,Nap}, disaggregated memory accessed through RDMA \cite{ziegler2019rdmaindex,Sherman,Deft,smart,DEX,CHIME,ROLEX,dLSM}, and DDR--CXL memory \cite{SIDLE}. Differences in addressing, access performance, and persistence motivate designs for node references, access structures across tiers, and the distribution of index components.

\noindent \textbf{Memory--Disk Systems.} Because disks do not support direct byte-addressable access, SQLite and PostgreSQL \cite{sqlite,postgresql_buffer} reference child nodes by logical page or block identifiers. A page cache or buffer manager resolves each identifier to a resident page; a miss triggers a disk read. References remain stable when pages enter or leave memory, but even memory hits require a page lookup. LeanStore \cite{LeanStore} uses pointer swizzling \cite{Swizzling} to reference resident pages directly through memory pointers and restores stable page identifiers on eviction. Anti-Caching \cite{anti} retains primary and secondary indexes in memory, locating tuples through direct pointers or entries in an evicted table that identify disk blocks and offsets. Its index memory footprint grows with data volume.

Other designs maintain separate access structures for data in different tiers. Siberia \cite{Siberia}, built on Hekaton \cite{Siberia2}, removes cold-record keys from in-memory indexes when moving records to persistent storage, using compact filters to avoid unnecessary cold-storage accesses. 2-Tree \cite{two} maintains hot and cold trees, accessing the cold tree after a hot-tree miss and migrating records between them based on access patterns observed at runtime. Tiered-Indexing \cite{Tiered-Indexing} extends this separation and record-level migration to B$^{+}$ trees, dynamic hash tables, indexed heaps, and log-structured merge (LSM) trees. Bf-Tree \cite{Bf-Tree} instead decouples disk leaf pages from their cached contents. Variable-sized mini-pages hold hot records and recent updates and can grow into full pages, supporting record-level caching, page-level caching, and update buffering as access patterns change.

\noindent \textbf{DRAM--Persistent Memory Systems.} Byte addressability and persistence motivate designs that organize pages and index components across DRAM and persistent memory. FOEDUS \cite{FOEDUS} maintains mutable DRAM pages and immutable persistent-memory snapshots, referencing both versions through dual-page pointers. Transactions update DRAM pages, while background threads merge logs with existing snapshots. FPTree \cite{FPTree} places internal nodes in DRAM and leaves in persistent memory, reconstructing internal nodes from persistent leaves during recovery. HiKV \cite{HiKV} combines a persistent-memory hash index for point queries with a DRAM-resident B$^{+}$ tree for range queries, avoiding B$^{+}$ tree maintenance in persistent memory.

DPTree \cite{DPTree} uses a DRAM buffer tree to absorb updates and merges them in batches into a base tree with DRAM-resident internal nodes and persistent-memory leaves. Point queries check the buffer before the base tree, while range queries combine results from both. BonsaiKV \cite{BonsaiKV} keeps its upper index and location information in DRAM, with persistent logs and data nodes in persistent memory. Aggregating multiple records in each data node reduces DRAM index space. For multi-socket systems, Nap \cite{Nap} adds a NUMA-aware layer over persistent-memory indexes, using local persistent-memory views for hot-item writes and a global DRAM view for hot-item reads to reduce remote persistent-memory accesses.

\noindent \textbf{RDMA-Based Disaggregated Memory.} One-sided RDMA operations allow direct remote-memory access, but index lookups often require multiple remote accesses. Ziegler et al. \cite{ziegler2019rdmaindex} compare three ways of distributing tree indexes: an independent tree per server, a single global tree distributed across servers, and a hybrid combining key-range partitioning with distributed leaf nodes. Sherman \cite{Sherman} modifies leaf layout and consistency checks to support partial writes and combines dependent RDMA operations to reduce round trips. Deft \cite{Deft} divides internal nodes into independently readable segments and uses hashing within leaves, avoiding full-node reads.

Compute nodes can also cache remote tree nodes in local DRAM to reduce remote accesses. SMART \cite{smart} stores a complete ART on memory nodes and caches some internal nodes on compute nodes. It validates cached references to detect invalidation caused by ART structure changes. DEX \cite{DEX} caches internal and leaf nodes along contiguous paths from the root. It partitions key ranges among compute nodes to reduce cache coherence overhead. On a miss, DEX uses estimated execution costs to choose between fetching remote nodes and offloading the remaining index operation to a memory node.

Other designs combine index structures or distribute data access and index maintenance tasks across compute and memory nodes. CHIME \cite{CHIME} combines B$^{+}$ tree routing with hashing within leaf nodes to reduce the memory required to cache the index and limit remote read amplification. ROLEX \cite{ROLEX} uses learned models to locate remote ordered data, with one-sided RDMA updates by compute nodes and asynchronous model retraining on memory nodes. dLSM \cite{dLSM} keeps write buffers on compute nodes, stores sorted tables in remote memory, and performs compaction on memory nodes to reduce cross-node data transfer.

Beyond optimizations for individual index structures, Hao et al. \cite{Hao} systematically compare five organizations for local memory, remote memory, and disk. These organizations differ in the number of B-trees and the storage tiers covered by each tree. Remote memory is covered by a tree spanning multiple tiers, used to cache disk-tree pages, or managed by a separate tree. Nodes are addressed through direct pointers or page-identifier mappings. Their analysis covers query processing, buffer management, and record migration.

\noindent \textbf{DDR--CXL Memory.} SIDLE \cite{SIDLE} exploits the shared DDR--CXL address space to manage index distribution at tree-node granularity, avoiding unnecessary migration associated with page-level management. It tracks only leaf access frequencies to identify hot paths. Upper-level and hot-path nodes are prioritized for DDR memory, and placement and migration respect parent--child relationships so that a root-to-leaf path crosses the tier boundary at most once.

Earlier designs address page resolution and disk access, asymmetric persistent-memory writes, or RDMA communication and data transfer. DDR and CXL memory both support direct CPU load/store access and have similar read/write characteristics, with a relatively limited latency and bandwidth gap. Results based on medium-specific overheads therefore do not directly characterize indexing trade-offs in this setting. Existing CXL indexing research remains limited. Our study systematically compares indexing designs in terms of request execution, data distribution adjustment, and DDR memory efficiency.

\section{Indexing Designs}\label{3}

\subsection{Design Overview}

\begin{figure*}[htbp]
\centering
\includegraphics[width=\textwidth]{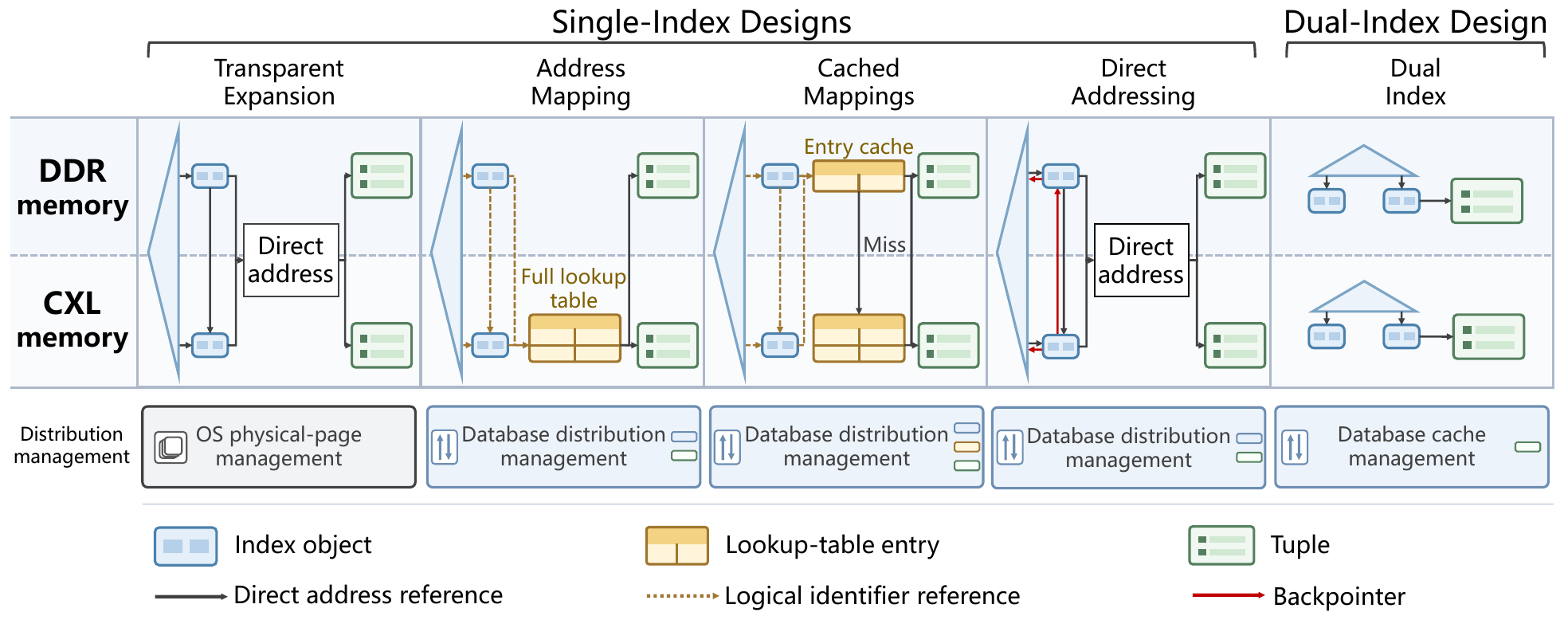}
\caption{Overall architecture of the five proposed indexing designs.}
\label{design}
\end{figure*}

Both DDR and CXL memory are directly addressable by the processor through the same load/store interface. A database can therefore treat them as a unified data space and use a single index to manage tuples distributed across both memory tiers. Alternatively, it can manage the two tiers separately and maintain an index for the tuples in each tier. These two forms of organization are referred to as single-index and dual-index designs, respectively. Figure~\ref{design} illustrates the five indexing designs proposed in this paper.


The single-index design with transparent memory expansion provides the simplest way to expand capacity using CXL memory without modifying the database implementation. The database does not explicitly distinguish the memory tiers in which indexes and tuples reside. Explicit distribution management, in contrast, allows the database to determine and adjust the distribution of indexes and tuples according to their access characteristics. This requires the single index to represent and resolve storage locations across the two tiers.

Conventional buffer management locates pages through logical page identifiers and a lookup table \cite{sqlite,postgresql_buffer}. The single-index design with address mapping applies the same indirect addressing principle: the index stores stable logical identifiers, and a lookup table maps these identifiers to memory addresses. Since an in-memory database manages data at tuple granularity, it requires tuple-level location mappings. As data volume grows, keeping the full lookup table in DDR memory would consume substantial space that could otherwise hold indexes and tuples. The full lookup table is therefore placed in CXL memory.

This placement can introduce an additional CXL memory access during address resolution. To reduce this overhead, the single-index design with address mapping is optimized in two ways. The single-index design with cached address mappings retains the original mapping and caches frequently accessed lookup-table entries in DDR memory to accelerate address resolution. The single-index design with direct addressing instead stores actual memory addresses in index entries, taking advantage of direct processor access to both tiers and eliminating the lookup table.

The dual-index design fixes one index in each memory tier and requires each index and the tuples it manages to reside in the same tier. For brevity, the five designs are referred to as Transparent Expansion, Address Mapping, Cached Mappings, Direct Addressing, and Dual Index, respectively. 

\subsection{Comparison Dimensions}

The indexing designs differ in their data access paths, data migration paths, and supported data distributions. These differences affect runtime performance, sensitivity to changes in CXL memory latency, and the performance benefit obtained from DDR memory. The designs also require different scopes and degrees of modification to the database. The comparison uses the following six dimensions to characterize these differences and trade-offs.

\textbf{Single-Record Operation Performance.} Lookups, inserts, updates, and deletes on individual tuples are basic operations supported by database indexes and common requests in database workloads. These operations involve locating tuples through the index and modifying index entries when tuples are inserted or deleted, or when indexed keys change. Different designs require lookups and updates in different indexes and incur different numbers of DDR and CXL memory accesses. These differences directly affect single-record operation performance.

\textbf{Range Query Performance.} Data-intensive applications, including data analytics, time-series processing, and real-time decision-making, often select tuples within a range of values on ordered attributes. A range query first locates the first index entry whose key is no smaller than the lower bound, then traverses subsequent entries in order and accesses their tuples until the key exceeds the upper bound. Different indexing designs incur different overheads during this traversal and tuple access. These overheads recur throughout the traversal, amplifying performance differences as the query range grows. Range queries therefore have different execution characteristics and indexing requirements from single-record operations.

\textbf{Data Migration Overhead.} Tuple access characteristics can change during execution, making the existing data distribution unsuitable for the current workload. Data migration adjusts this distribution by copying data across tiers while ensuring that foreground requests can correctly locate the migrated objects during and after migration. The designs differ in the objects that must be moved and the location information or index entries that must be updated after a location change. They consequently incur different data transfer, memory access, and synchronization overheads. Since DDR and CXL memory have relatively similar access performance and both provide low access latency, the performance benefit of distribution adjustments is relatively limited. Migration overhead can offset part of this benefit and affect both overall performance and the ability to adapt quickly to changes in tuple access characteristics.

\textbf{Robustness to CXL Memory Latency.} CXL memory devices can differ in their memory technologies and controller implementations, as well as in link configurations and connection topologies. Their actual access latency is therefore not fixed. The data access and migration paths of different designs involve different numbers of CXL memory accesses. Higher CXL memory latency amplifies the associated overheads to different degrees, changing the performance differences between designs and their suitability for different devices and topologies.

\textbf{DDR Memory Efficiency.} DDR memory is the fast, capacity-limited tier, and all five designs use its performance advantage to improve database performance. However, they use DDR memory in different ways, storing different indexes, tuples, and address information. These contents have different memory footprints and different effects on data access and migration paths. The designs can therefore achieve different performance with the same DDR capacity, reflecting different DDR memory efficiency.

\textbf{System Integration Complexity.} CXL.mem makes expanded memory directly accessible to the processor, allowing transparent capacity expansion without changing existing database data structures or request execution. Such expansion, however, does not allow the database to control distribution explicitly according to tuple access characteristics. Explicit control requires changes to index structures and address resolution mechanisms, with corresponding changes to index lookups, updates, and data migration. Broader and more extensive changes generally increase implementation difficulty and the degree of intrusion into the database. Stronger dependence on a particular database implementation also requires more changes when applying the design to other systems, reducing generality and portability. Since integration complexity cannot be captured by a single numerical measure, the database components affected and the extent of their modification provide the basis for qualitative comparison.

\subsection{Transparent Expansion}
The single-index design with transparent memory expansion uses CXL memory to expand capacity. CXL memory is exposed as a separate NUMA node under operating system memory management, allowing the database to retain its existing indexing design and request execution. The operating system handles memory allocation. Under the default NUMA allocation policy, allocations first use local DDR memory and fall back to CXL memory when DDR capacity is insufficient. The initial distribution of indexes and tuples is therefore determined primarily by allocation order and available capacity at allocation time.

The database does not explicitly manage the tiers in which indexes and tuples reside. During execution, their locations are adjusted only through operating system page migration across tiers, which is transparent to the database. Since operating system hotness tracking and migration do not operate at tuple or index-node granularity, the indexes and tuples resident in DDR memory may not correspond to the current hotspots. This design expands the amount of data the database can manage without requiring implementation changes. Its ability to exploit DDR performance nevertheless depends on how well the resulting distribution matches the actual access characteristics.

\subsection{Address Mapping}\label{34}

To control data distribution explicitly, the Address Mapping design stores stable logical identifiers for tuples in the index and maintains a lookup table mapping these identifiers to tuple addresses. This follows the indirect addressing used in conventional buffer management. In a memory-disk system, disks do not support direct byte-addressable processor access, so the full lookup table resides in DDR memory to avoid additional disk accesses during address resolution. Page-level mappings also have a relatively small metadata memory footprint. In a hybrid DDR-CXL database, however, tuple-level data management requires tuple-level mappings. The full lookup table grows with data volume, and storing it entirely in DDR memory would consume substantial space and may even exhaust the available capacity. Since CXL memory is directly addressable by the processor, the full table is placed in CXL memory, leaving more DDR space for frequently accessed indexes and tuples.

When a tuple migrates between DDR and CXL memory, its logical identifier remains unchanged. The database only needs to copy the tuple and update its address in the corresponding lookup-table entry. The original index entry remains unchanged and still locates the tuple through the same identifier, simplifying location maintenance during migration. However, this simplification adds address resolution to tuple access. After an index lookup returns a tuple identifier, the database must query the lookup table to obtain the actual address. Since the full table resides in CXL memory, address resolution can incur an additional CXL access even when the tuple itself resides in DDR memory. When multiple tuples are accessed in sequence, this resolution is repeated and its overhead accumulates with the number of tuples. Higher CXL memory latency further increases the mapping access overhead.

Implementing this design requires changes to the memory allocator and index-entry representation, together with an interface through which data accesses interact with an independent policy layer. The allocator separately manages available space in DDR and CXL memory and supports allocation and deallocation in a specified memory tier \cite{nephele}. Tuple addresses in index entries are replaced by stable logical identifiers. Through the interface, the policy layer tracks hotness and determines and adjusts data distribution. It also queries and maintains the full lookup table, returning the tuple after resolving its address.

\subsection{Cached Mappings}

To reduce the CXL memory accesses introduced by address resolution, the Cached Mappings design extends Address Mapping by caching frequently accessed table entries in DDR memory. Index lookup and traversal remain unchanged; the cache is consulted after the logical identifier is obtained. On a cache hit, address resolution completes in DDR memory, reducing CXL accesses. On a miss, the database still queries the full table in CXL memory, adding a cache lookup compared with the Address Mapping design. Cached entries also consume DDR space and compete with frequently accessed indexes and tuples for the limited capacity.

During tuple migration, the database copies the tuple across tiers and updates its address in the corresponding table entry. If the entry is cached in DDR memory, the database updates the cached entry directly and writes it back to the full table in CXL memory when it is evicted. If the entry is not cached, the policy layer decides whether to admit it and updates the entry accordingly.

Implementation builds on the changes described in Section \ref{34}. The independent policy layer is extended to incorporate the entry cache into address resolution and manage the cache state of table entries according to their accesses during execution.

\subsection{Direct Addressing}

The single-index design with direct addressing explicitly controls distribution while avoiding accesses to the full lookup table during address resolution. Since the processor can directly address both memory tiers, this design stores actual tuple addresses in index entries. The database retains its original index lookup and traversal procedures and accesses tuples directly through these addresses. Neither the full lookup table nor its entry cache is required.

Direct addresses eliminate address resolution during tuple access, but migration requires more than copying the tuple. All pointers to the tuple must be updated to its new address so that subsequent requests locate it correctly. The database cannot directly identify where these pointers are stored, and traversing the index again to update them would incur substantial migration overhead. To avoid this traversal, the design maintains a set of backpointers for each tuple, recording the locations of all pointers that refer to it. These backpointers consume additional memory and must be updated when data migrates or the index structure changes, adding migration and index maintenance overhead.

This design reuses the tier-aware allocator, policy-layer interface, and distribution management described in Section \ref{34}. It additionally modifies index-entry representation and maintenance to incorporate backpointers and update them during data migration and index maintenance.

\subsection{Dual Index}

The dual-index design maintains separate indexes in DDR and CXL memory, each managing tuples in its own tier. The CXL index manages all tuples, whereas the DDR index manages copies of frequently accessed tuples. Requests for tuples covered by the DDR index can locate and access their DDR-resident copies directly through that index. Hot-tuple accesses thus bypass the full CXL index, shortening the index lookup path and reducing access overhead. For other tuples, a lookup that misses in the DDR index must continue in the full CXL index, adding an index lookup. The DDR index and its tuple copies also occupy the limited DDR capacity.

Since all tuples and the full index remain in CXL memory, range queries can traverse the CXL index directly. However, the latest versions of some tuples reside in DDR memory. The database must also query the DDR index and perform additional data synchronization to return the latest version of each tuple. Distribution adjustments create or remove DDR-resident tuple copies together with their index entries. Modified copies must be written back to CXL memory before removal.

The independent policy layer manages the DDR index and its tuple copies. The design reuses the allocator, policy-layer interface, and distribution management described earlier.

\subsection{Design Trade-offs}


Table~\ref{tab:indexing-overheads} summarizes the additional operations and overheads introduced by the five designs in data access, data migration, memory access and consumption, and database-internal modifications. "None" indicates that the indexing design introduces no additional operation or resource overhead in that category. Transparent expansion requires no database modifications, but the database cannot actively control distribution according to access characteristics. The other four designs all require a tier-aware allocator and an interface between data access and the independent policy layer. The policy layer handles hotness tracking, distribution management, and the queries and structure maintenance required by each design. Since these tasks are handled by the policy layer, additional runtime operations do not directly correspond to the scope of database-internal modifications. The last row lists only the database-internal changes beyond the common allocator and interface modifications.

\begin{table}[t]
\centering
\caption{Additional operations and overheads of the five indexing designs.}
\label{tab:indexing-overheads}
\begingroup
\footnotesize
\setlength{\tabcolsep}{3pt}
\renewcommand{\arraystretch}{1.2}
\renewcommand{\tabularxcolumn}[1]{m{#1}}
\begin{tabularx}{\linewidth}{
    @{}
    >{\raggedright\arraybackslash}m{0.14\linewidth}
    >{\centering\arraybackslash}X
    >{\hsize=0.9\hsize\linewidth=\hsize\centering\arraybackslash}X
    >{\hsize=1.1\hsize\linewidth=\hsize\centering\arraybackslash}X
    >{\centering\arraybackslash}X
    >{\centering\arraybackslash}X
    @{}
}
\toprule
\textbf{Aspect}
& \textbf{Transparent}
& \textbf{Address Mapping}
& \textbf{Cached Mappings}
& \textbf{Direct Addressing}
& \textbf{Dual Index} \\
\midrule
Single-record access
& None
& Table lookup
& Cache lookup
& None
& DDR index lookup \\
Range query
& None
& Table lookup
& Cache lookup
& None
& Tuple-version sync. \\
Migration
& None
& Mapping updates
& Updates; writeback
& Backpointer updates
& DDR index updates \\
CXL accesses
& None
& Table lookup
& Uncached mappings
& None
& None \\
DDR space
& None
& None
& Cached mappings
& Backpointers
& DDR index \\
Database changes
& None
& Index entries
& Index entries
& Index entries and maintenance
& No further changes \\
\bottomrule
\end{tabularx}
\par\smallskip
\begin{minipage}{\linewidth}
\raggedright
For range queries, the listed operations apply to each tuple.
Cached Mappings queries the lookup table on cache misses.
\end{minipage}
\endgroup
\end{table}

Explicit distribution control requires the database to represent and resolve tuple locations across the two tiers. The Address Mapping design uses stable identifiers and a lookup table, so migration only requires updating the corresponding table entry, but every tuple access requires an additional lookup in CXL memory. The Direct Addressing design eliminates this resolution step and retains the original tuple access procedure. It instead requires backpointers and their maintenance during migration and index structure changes, increasing the overheads of these operations. The trade-off therefore depends on the relative frequencies of tuple access, data migration, and index structure changes.

Caching frequently accessed CXL-resident contents in DDR memory reduces CXL accesses on hot paths. Cached lookup-table entries and the DDR index accelerate address resolution and tuple location, respectively. These structures occupy DDR space that could otherwise hold indexes or tuples, and misses incur additional lookups. Such designs trade DDR space for faster hot-data access. Their actual benefit depends on request type, access skew, and CXL memory access overhead.
\section{Adaptive Data Distribution Optimization}\label{4}

\subsection{Data Distribution Constraints}

Indexing design and data distribution jointly affect database performance. Data distribution must therefore be determined and adjusted according to access characteristics. Different indexing designs impose different constraints on distribution management. These constraints concern whether the database explicitly manages the distribution of indexes and tuples across the two memory tiers, whether indexes and tuples can be placed separately, and which objects are involved in distribution adjustments.

Dual Index stores all tuples and the complete index in CXL memory. DDR memory holds copies of frequently accessed tuples together with a separate index for these copies. Both indexes remain fixed in their respective memory tiers, and each index manages tuples stored in the same tier. When a tuple copy is created or removed in DDR memory, the corresponding entry is inserted into or deleted from the DDR index. Distribution management therefore operates on tuples and only needs to determine which tuples retain copies in DDR memory. For this tuple caching task, Dual Index directly adopts the lightweight, benefit-aware caching policy of Nephele \cite{nephele}.

\begin{figure*}[htbp]
\centering
\includegraphics[width=\textwidth]{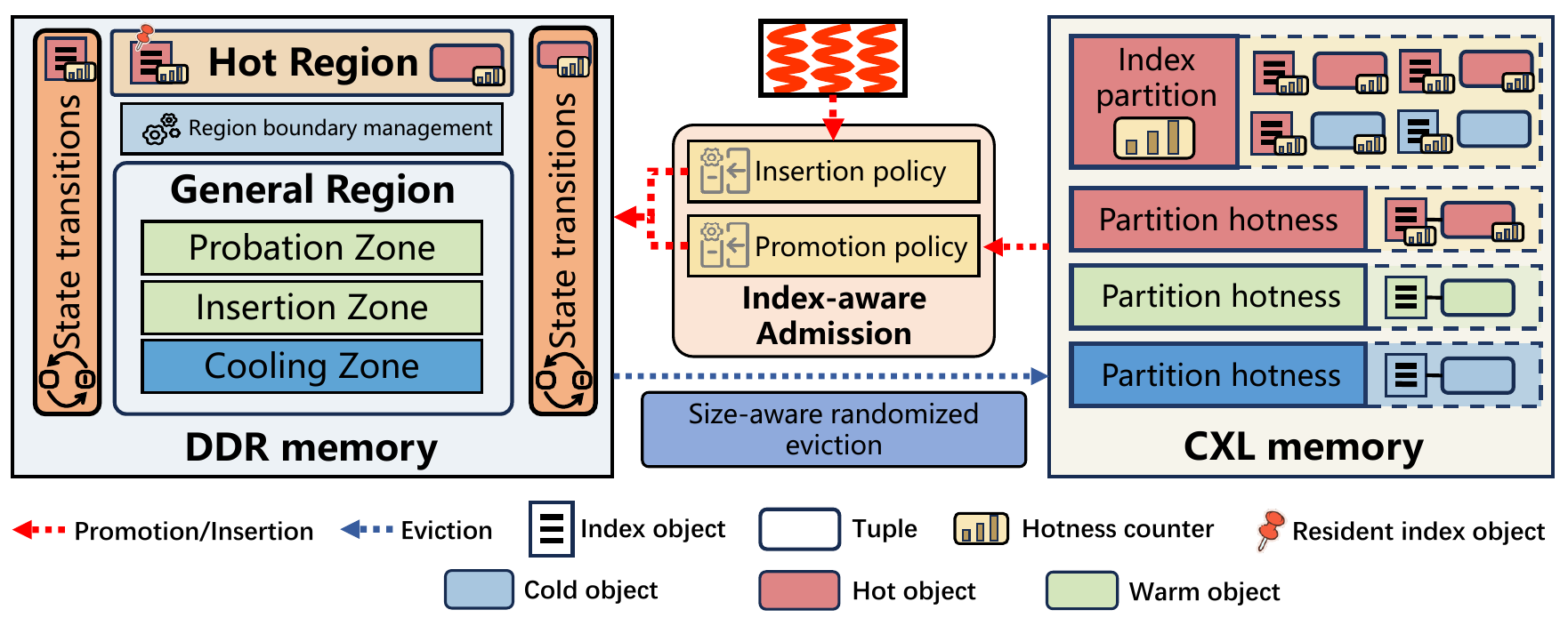}
\caption{Overview of separate management of index objects and tuples.}
\label{overview}
\end{figure*}

Transparent Expansion relies on the operating system to determine the initial data distribution and adjust it at runtime. The database is unaware of the memory tiers in which indexes and tuples reside and does not explicitly determine or adjust their distribution. In the other three single-index designs, the database explicitly manages data distribution, and one index manages all tuples across both tiers. If distribution adjustments operate on complete index nodes, the index portion and associated tuples must reside in the same memory tier. However, the index portion and tuples may become hot at different times. Moving an entire node to DDR memory can therefore waste capacity on components that are not hot and reduce performance. To address this limitation, the portion of an index node used to locate and match tuples is defined as an index object. The proposed method extends Nephele to determine and adjust the distributions of index objects and tuples separately \cite{nephele}. Figure~\ref{overview} provides an overview of the method.


\subsection{Index-aware Admission and Eviction}
\label{sec:index-aware-admission-eviction}

DDR memory is divided into a Hot Region and a General Region to support the admission and eviction of index objects and tuples. Objects with validated high access frequency are retained in the Hot Region. The General Region accommodates objects that have not yet been confirmed as hot but may become hot subsequently. It contains three zones. The Insertion Zone holds newly created index objects and newly inserted tuples, the Probation Zone holds objects promoted from CXL memory, and the Cooling Zone holds objects whose hotness has declined.

The method adopts the adaptive region boundary management mechanism of Nephele \cite{nephele}. During the initial execution period, the boundary between the Hot Region and the General Region is gradually determined according to hotspot growth to balance performance and DDR memory utilization. A substantial and sustained decline in throughput triggers renewed hotspot detection and boundary adjustment to adapt to changes in access patterns.

\noindent \textbf{Benefit-aware admission.}
Index objects and tuples both use epoch-based hotness estimation to provide access statistics for benefit-aware admission. The estimator uses globally consistent logical timestamps, called epochs, to measure relative intervals between accesses. These timestamps advance monotonically according to the number of accesses processed by the database. Table~\ref{tab:distribution-symbols} summarizes the main parameters used in this section.

\begin{table}
  \caption{Parameter Description}
  \label{tab:distribution-symbols}
  \vspace{-1mm}
  \begin{tabular}{cl}
    \toprule
    Parameters & Description \\
    \midrule
    $ref$ & Hotness value of an index object or tuple \\
    $\gamma$ & Decay rate ($0 < \gamma < 1$, default: 0.5) \\
    $epoch_{now}$ & Current epoch \\
    $epoch_{last}$ & Epoch of the most recent access \\
    $\alpha_{\text{sat}}$ & Saturation parameter (default: 8) \\
    $N_p^{\mathrm{idx}}, N_p^{\mathrm{tuple}}$
      & Admission hotness thresholds for index objects and tuples \\
    $N_{bp}$ & Threshold to enable index object hotness tracking \\
    $c \in \{\mathrm{idx},\mathrm{tuple}\}$
      & Object class (index object or tuple) \\
    $[\alpha_{I}^c,\beta_{I}^c,\alpha_{P}^c,\beta_{P}^c]$
      & Coefficients for predicting transition probabilities \\
    $P_{i}^c$, $\hat{P}_{i}^c$
      & Obs./Pred. transition prob. (Probation $\to$ Hot Region) \\
    $I_{i}^c$, $\hat{I}_{i}^c$
      & Obs./Pred. transition prob. (Insertion $\to$ Hot Region) \\
    \bottomrule
  \end{tabular}
  \vspace{-5mm}
\end{table}

For each object, the estimator records a hotness value $ref$ and the epoch of the most recent access, $epoch_{last}$. An access in epoch $epoch_{now}$ updates hotness as:
\[
ref = \min(\alpha_{\text{sat}} \cdot N_p, ref_{old} \cdot \gamma^{(epoch_{now} - epoch_{last})} + 1)
\]
where $ref_{old}$ is the hotness value before the update, $\gamma$ is the decay rate, $\alpha_{\text{sat}}$ is the saturation parameter, and $N_p$ is the admission hotness threshold for the object. The update combines accumulated access counts with exponential decay based on the interval between accesses. Hotness thus reflects both access frequency and recency, while decay reduces the influence of outdated access history. The saturation bound $\alpha_{\text{sat}} \cdot N_p$ prevents unbounded growth and keeps decay effective after many accesses. The hotness value and the epoch of the most recent access are updated only when the object is accessed, avoiding global updates to these values at every epoch. This yields a unified hotness estimator that produces comparable estimates with low maintenance overhead.

Admission is handled separately by the promotion and insertion policies. The promotion policy uses $N_p$ to assess the expected benefit of moving an object from CXL to DDR memory. Promotion incurs migration overhead, including data copying between the tiers. Each subsequent access served from DDR rather than CXL reduces latency by the difference between the two tiers. The threshold $N_p$ therefore represents the minimum number of such accesses needed to offset one promotion. An object with $ref > N_p$ is expected to yield a positive benefit and is promoted to the Hot Region. An object with $\gamma \cdot N_p < ref \leq N_p$ may become hot and enters the Probation Zone. Newly inserted tuples and newly created index objects have no reliable hotness history, and their initial allocation requires no migration between tiers. The insertion policy therefore initially admits both to the Insertion Zone. Observations from the two validation zones subsequently guide adjustments to the respective admission probabilities.

Index objects and tuples differ in memory footprint, benefit per access, and the overhead of maintaining location information. They consequently require different numbers of accesses to offset their promotion overhead. Equal hotness values indicate similar access frequency and recency, but do not justify using one threshold to assess the promotion benefit of both. The admission thresholds are therefore denoted separately by $N_p^{\mathrm{idx}}$ for index objects and $N_p^{\mathrm{tuple}}$ for tuples. Each threshold is determined by the promotion overhead and benefit per access of the corresponding object class. The two classes use their respective thresholds to determine admission to the Hot Region or Probation Zone and to set the saturation bound together with $\alpha_{\text{sat}}$. This allows the same hotness estimator to support separate admission decisions for index objects and tuples.

\noindent \textbf{Adaptive admission.}
The Insertion and Probation Zones accommodate potential hotspots to improve DDR memory utilization. If admitted objects are not accessed frequently enough, however, they cause cache pollution and reduce performance. The admission policy therefore adjusts subsequent admission probabilities based on the observed transitions of admitted objects into the Hot Region, so that DDR capacity is preferentially allocated to objects more likely to become hot.

An index object may participate in locating and matching multiple tuples. Accesses to these tuples collectively contribute to the hotness of the index object, whereas tuple hotness results directly from accesses to that tuple. The two object classes can therefore exhibit different hotness trends. Their transition probabilities from the General Region to the Hot Region are estimated independently.

Algorithm~\ref{alg:admission-update} describes the adjustment procedure. At the end of each epoch, the policy records, separately for index objects and tuples, the fractions of objects that transition from the Insertion and Probation Zones to the Hot Region. These fractions are the observed transition probabilities for that epoch (line 5). Lightweight least-squares fitting uses observations accumulated across successive epochs to estimate the trends under the current access pattern. The fitted results are combined with the predictions from the previous epoch to produce insertion and promotion admission probabilities for the next epoch (lines 6--9). The update adapts admission to changes in access patterns while reducing the influence of fluctuations within an individual epoch.

\begin{algorithm}
\caption{Adaptive update of admission probabilities.}
\label{alg:admission-update}

\nonl\textbf{fitting(x)} :
     Re-estimates the corresponding predictor coefficients using least squares.\\

\LinesNumbered

\nonl$\alpha_{I}^c, \beta_{I}^c, \alpha_{P}^c, \beta_{P}^c \gets \textsc{Rand}(0,1)$ for each $c \in \{\mathrm{idx},\mathrm{tuple}\}$.\\

\While{true}{
    \If{a new epoch begins}{
        $i \gets epoch_{now} - 1$\\
        \ForEach{$c \in \{\mathrm{idx},\mathrm{tuple}\}$}{
            $P_{i}^c = n_{i,c}^{prob\_hot}/n_{i,c}^{prob}$; $I_{i}^c = n_{i,c}^{insert\_hot}/n_{i,c}^{insert}$\\
            $\alpha_{I}^c, \beta_{I}^c \gets fitting(I_{i}^c, \hat{I}_{i}^c)$\\
            $\alpha_{P}^c, \beta_{P}^c \gets fitting(P_{i}^c, \hat{P}_{i}^c)$\\
            $\hat{I}_{i+1}^c \gets \alpha_{I}^c \times I_{i}^c + \beta_{I}^c \times \hat{I}_{i}^c$\\
            $\hat{P}_{i+1}^c \gets \alpha_{P}^c \times P_{i}^c + \beta_{P}^c \times \hat{P}_{i}^c$\\
        }
    }
}
\end{algorithm}

\noindent \textbf{Access-driven data state transitions.}
The state transition mechanism manages index objects and tuples across the Hot Region, Insertion Zone, Probation Zone, and Cooling Zone according to their expected benefit. Transitions are implemented by updating the metadata tag of each object. Objects with declining hotness gradually move to zones of lower priority and are eventually reclaimed by priority-guided randomized eviction. This preserves hot objects with low metadata overhead. The following rules are stated for index objects with threshold $N_p^{\mathrm{idx}}$. Tuples follow the same rules with threshold $N_p^{\mathrm{tuple}}$.

\noindent \textsc{Insertion and Probation Zones} $\to$ \textsc{Hot Region or Cooling Zone}. Index objects in either validation zone receive a protection window to prevent eviction before hotness validation is complete. An object enters the Hot Region if $ref > N_p^{\mathrm{idx}}$ within this window. After the window ends, an object whose hotness falls to $\gamma \cdot N_p^{\mathrm{idx}}$ or below enters the Cooling Zone.

\noindent \textsc{Hot Region} $\to$ \textsc{Cooling Zone}. An index object enters the Cooling Zone when decay across successive epochs reduces its hotness to $\gamma \cdot N_p^{\mathrm{idx}}$ or below.

\noindent \textsc{Cooling Zone} $\to$ \textsc{Hot Region, Probation Zone, or eviction}. The policy rechecks the current hotness of an index object selected from the Cooling Zone. If $ref > N_p^{\mathrm{idx}}$, the object enters the Hot Region. If $\gamma \cdot N_p^{\mathrm{idx}} < ref \leq N_p^{\mathrm{idx}}$, it enters the Probation Zone. Otherwise, it is evicted from DDR memory.

\noindent \textbf{Priority-guided randomized eviction.}
Index objects and tuples share the limited DDR capacity. When capacity becomes constrained, the policy reclaims space according to zone priority, starting with the Cooling Zone. If no reclaimable objects remain there, it selects objects from the Probation and Insertion Zones, prioritizing the zone with a lower probability of transition to the Hot Region. Objects within a zone have the same eviction priority but can differ substantially in memory footprint. Candidate selection therefore also accounts for the actual memory footprint of each object.

For each eviction, the policy selects a zone by priority and randomly samples at most three candidate objects. It rechecks their current hotness and evicts the object with the largest memory footprint among those that still satisfy the eviction conditions. If no reclaimable objects remain outside the Hot Region, newly created index objects and newly inserted tuples are placed directly in CXL memory, and further promotions are suspended for the current epoch. This avoids performance fluctuations caused by repeated admission followed by immediate eviction. The procedure uses only local information from a few candidates and requires no global metadata structure ordered by hotness or size. It preserves hot objects and improves DDR memory efficiency with low metadata management and synchronization overhead.

\subsection{Hierarchical Hotness Tracking}
\label{sec:hierarchical-hotness-tracking}

Both index objects and tuples require access hotness information to determine and adjust their distributions. Maintaining fine-grained statistics for all objects would substantially increase the metadata memory footprint and statistics update overhead. Hierarchical hotness tracking controls these costs by dividing index objects into disjoint logical partitions and initially maintaining coarse-grained hotness at the partition level. Each partition uses the epoch-based estimator described in Section~\ref{sec:index-aware-admission-eviction}.

A partition is marked as active when its hotness exceeds $N_{bp}$, enabling index object hotness tracking within that partition. Index object hotness uses the same epochs and decay semantics as partition hotness to maintain consistency between the two granularities. If no hot index objects are detected in an active partition for $\log_{1/\gamma}(\alpha_{\text{sat}})+1$ consecutive epochs, the partition is marked as cold and its index object metadata is reclaimed to reduce maintenance overhead.

Logical partitions are also split or merged according to their estimated hotness as access patterns change. To improve tracking precision, the policy splits a partition when its hotness persistently exceeds $\frac{N_{bp}}{\gamma^2}$, isolating hotspots. To reduce tracking overhead, it merges the partition with adjacent partitions when hotness remains below $\gamma \cdot N_{bp}$. These adjustments adapt the partitions to changes in hotspot distribution while preserving tracking accuracy and limiting metadata overhead.

Index object hotness further identifies which tuples require fine-grained tracking. An access to a tuple also updates the access statistics of the associated index object. Tuple hotness tracking is enabled only when the hotness of that index object exceeds $\min(N_p^{\mathrm{idx}},N_p^{\mathrm{tuple}})$, allowing the policy to determine whether the tuple can satisfy the admission conditions. When index object hotness falls to this threshold or below, tuple tracking stops and the corresponding metadata is reclaimed. Tracking therefore proceeds through logical partitions, index objects, and tuples. Fine-grained statistics are confined to objects that may qualify for admission, reducing both the metadata memory footprint and statistics update overhead.

\subsection{Index Object Residency Policy}
\label{sec:index-object-residency}

\begin{figure*}[htbp]
\centering
\includegraphics[width=0.8\textwidth]{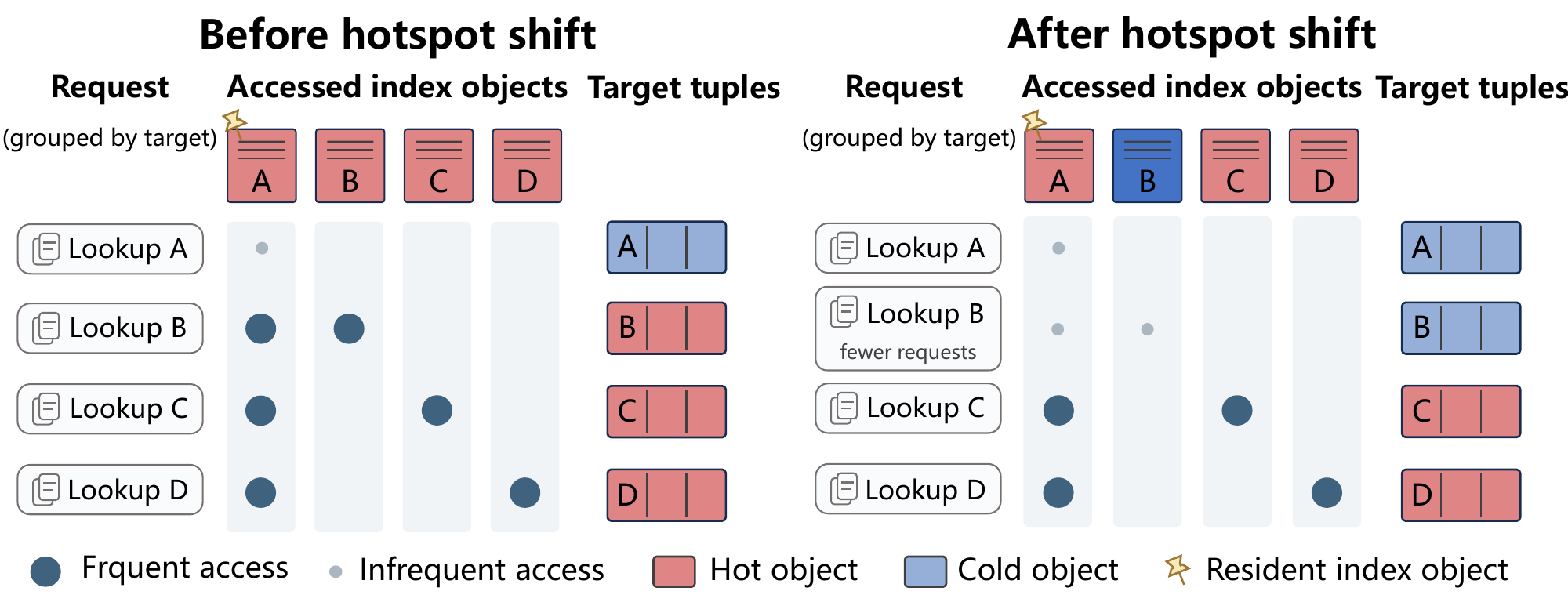}
\caption{Hotness sources and residency of index objects.}
\label{residency}
\end{figure*}

Hierarchical tracking reduces the number of index objects and tuples that require fine-grained statistics. A request may still access several index objects in sequence while locating and matching the target tuple, resulting in multiple hotness updates and substantial metadata management overhead. Figure \ref{residency} illustrates the two sources of index object hotness. An index object may lie on the lookup paths of multiple tuples, with accesses along these paths collectively contributing to its hotness. Upper-level nodes in tree indexes typically exhibit this behavior. Accesses to a particular tuple also contribute to the hotness of the associated index object. The residency policy uses the different hotness characteristics produced by these accesses to retain index objects that are more likely to remain hot in DDR memory.

When the hotness of an index object reaches the saturation bound $\alpha_{\text{sat}} \cdot N_p^{\mathrm{idx}}$, the policy checks whether the associated tuple satisfies the admission conditions to identify the main source of index object hotness. If the tuple qualifies, the hotness of the index object may depend on the current access pattern of that tuple. The policy therefore continues to track index object hotness and adjust the memory tier in which the object resides using the preceding mechanisms. If the tuple does not qualify, lookups for other tuples are more likely to be the main source of index object hotness. Because these accesses involve multiple tuples, the hotness of the index object is less sensitive to changes in the access pattern of any single tuple. The policy therefore retains that index object in DDR memory. Hotness information is no longer maintained for the resident index object, while the associated tuple continues to be managed according to its own hotness. This reduces the number of index objects requiring continuous tracking and distribution management, lowers statistics update and metadata management overhead, and avoids unnecessary adjustments between memory tiers.

\section{Experimental Evaluation}
\label{sec:evaluation}

\subsection{Experimental Setup}
\label{sec:experimental-setup}

\textbf{Machine Configuration.}
Experiments are conducted on a server equipped with two Montage Jintide\textregistered{} C8458P processors, 128~GB of DDR memory (four 32-GB modules), and 256~GB of CXL memory (four 64-GB modules). The server runs Ubuntu 24.04.3 LTS. To avoid NUMA effects, all experiments run on a single processor and use only the DDR and CXL memory connected to that processor, consistent with prior work \cite{nomad}. The available DDR and CXL capacities in this configuration are 64~GB and 128~GB, respectively. CXL memory is connected through the Montage Technology M88MX5891 CXL Memory Expander Controller. Memory access latencies are measured using Intel Memory Latency Checker (MLC) \cite{mlc}, yielding 115~ns for local DDR memory and 287~ns for CXL memory. These measurements are consistent with prior studies \cite{ddr2,ev1,ev2,nomad}, indicating that the platform reflects typical access characteristics of existing CXL memory devices and thereby supporting the generality of the experimental findings. CXL memory capacity is configured according to the database size, while DDR memory capacity is adjusted through $P$, defined as the ratio of DDR memory capacity to CXL memory capacity. The default value is $P=0.1$, consistent with prior work \cite{fifo}. The impact of other $P$ values is evaluated in Section~\ref{sec:ddr-capacity}.

\noindent \textbf{Benchmarks.}
The evaluation uses YCSB \cite{ycsb} to model data accesses in large-scale online services. The database contains a 32-GB relational table. Each tuple consists of an 8-byte primary key and ten columns, each holding a 100-byte random string. The workload configurations vary in read/write ratio and access skew. The percentages of read operations are 100\%, 90\%, 50\%, 10\%, and 0\%, with the remaining requests being writes. These configurations correspond to Read-Only (RO), Read-Heavy (RH), Balanced (BA), Write-Heavy (WH), and Write-Only (WO), respectively. Tuple accesses follow a Zipfian distribution, with $z$ set to 0.3, 0.6, and 0.9.

\noindent \textbf{Compared Designs and Implementation.}
A hybrid DDR--CXL database prototype was developed by extending DBx1000, an in-memory database research prototype \cite{dbx1000-2,dbx1000-1,zen,dbx1000-3,dbx1000-4}. It implements the five indexing designs presented in Section~\ref{3}, including the four single-index designs Transparent Expansion, Address Mapping, Cached Mappings, and Direct Addressing, as well as Dual Index. A skiplist is used as the underlying index structure to support range queries. Its maximum height is set to 12, consistent with existing implementations \cite{cache_shard_1}. The number of pointers in each index object is determined by its actual height, and each pointer occupies 8~bytes. Data distribution in Transparent Expansion is managed by the operating system. Dual Index uses Nephele \cite{nephele} to manage tuple copies in DDR memory. The other three single-index designs use the data distribution management method described in Section~\ref{4}, which manages index objects and tuples separately.

Requests are executed by 16 threads. To mitigate lock contention, each data distribution policy implemented in the database uses 32 shards, with an independent instance for each shard, consistent with common implementations in existing systems \cite{cache_shard_1,cache_shard_2}. Whenever the processed access count reaches five times the number of tuples that DDR memory can hold, the global epoch advances, consistent with prior research and production practice \cite{TinyLFU,decay_1}. Address Mapping, Cached Mappings, and Direct Addressing compute separate admission hotness thresholds, $N_p^{idx}$ and $N_p^{tuple}$, for index objects and tuples, whereas Dual Index computes only the tuple admission hotness threshold. The thresholds are recomputed at intervals of 10 epochs using the average access latencies of DDR and CXL memory and the migration overhead of the corresponding object type. For the four designs in which the database determines the memory tier of an object, the managed objects are initially allocated in CXL memory. Transparent Expansion uses the default NUMA allocation policy. All performance measurements are collected after warm-up, once the database reaches a steady state.

\subsection{Overall YCSB Performance}
\label{sec:overall-performance}

\begin{figure*}[t]
\centering
\includegraphics[width=\textwidth]{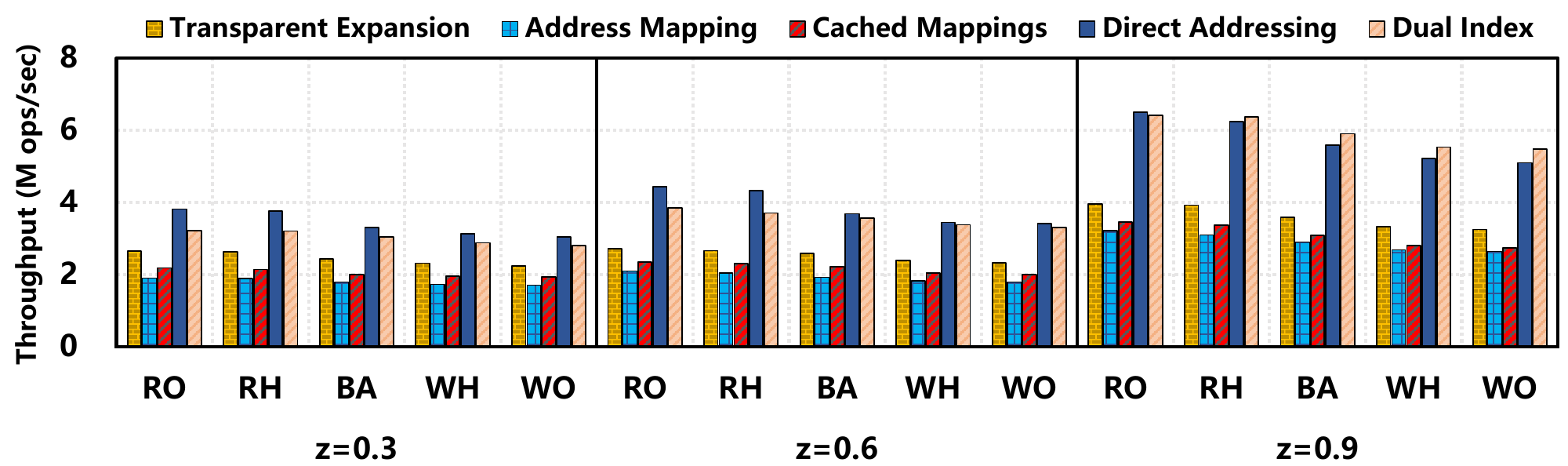}
\caption{Overall throughput under YCSB.}
\label{overall-5}
\end{figure*}

YCSB workloads with different combinations of read/write ratios and values of the Zipfian parameter $z$ are used to evaluate the overall performance of the five indexing designs. The main results are shown in Figure \ref{overall-5}.

Address Mapping stores stable logical identifiers in the index and maintains their mappings to object addresses in a lookup table. This allows the database to explicitly control the distribution of index objects and tuples between DDR and CXL memory. However, its throughput is only 71.64\%--81.49\% of that of Transparent Expansion across all evaluated workloads, even though Transparent Expansion does not account for workload characteristics. This is because accessing an object along the index lookup path first requires a lookup-table query to resolve its actual memory address, adding one table lookup to each object access. Since the full lookup table resides in CXL memory, these queries introduce additional CXL memory accesses into the index lookup path.

\begin{figure}[t]
  \centering
  
  \begin{minipage}{0.48\textwidth}
    \centering
    \includegraphics[width=\linewidth]{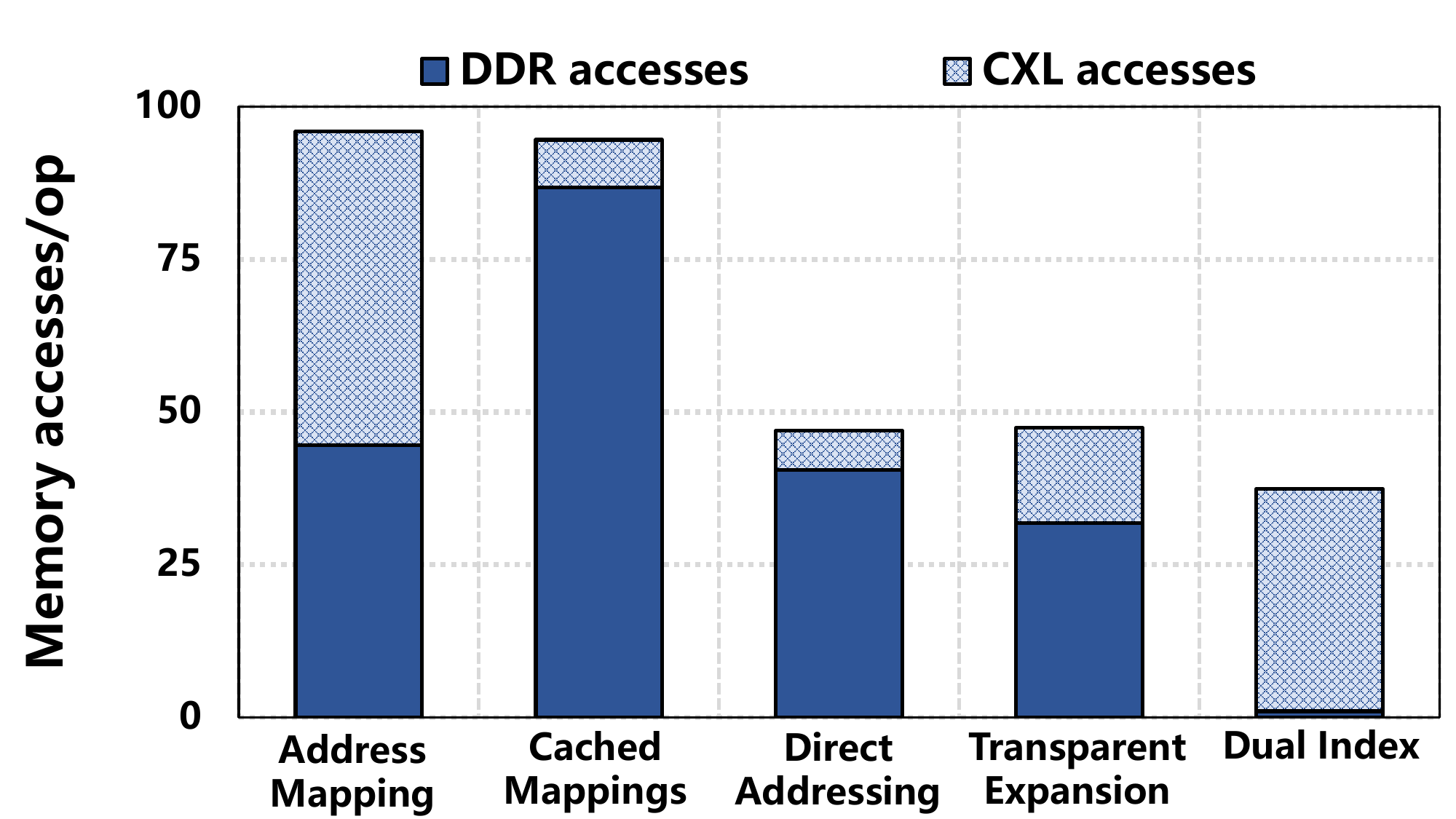}
    \caption{Memory accesses per request under YCSB.}
    \label{accesses-5}
  \end{minipage}%
  \hfill
  \begin{minipage}{0.48\textwidth}
    \centering
    \includegraphics[width=\linewidth]{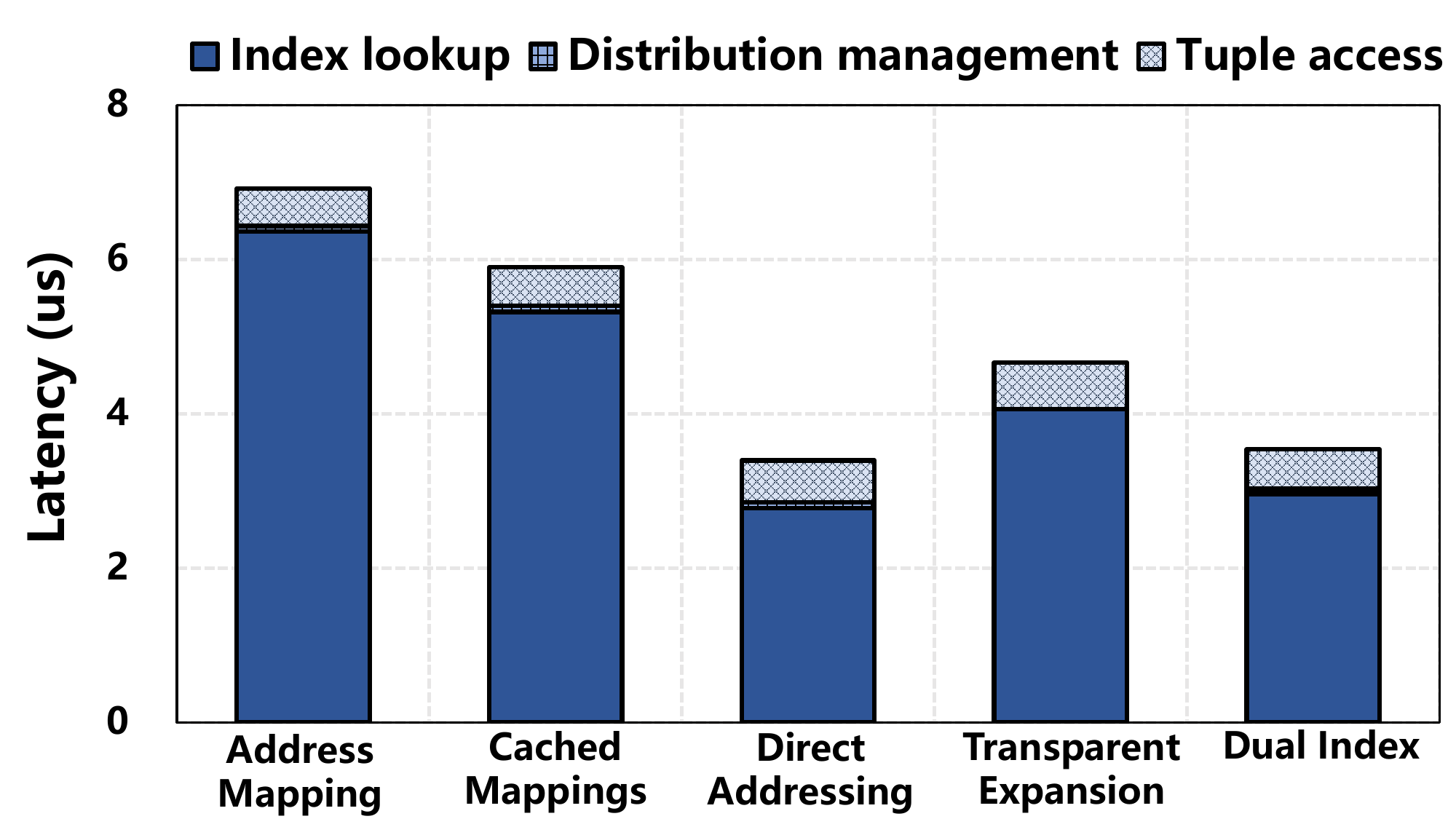}
    \caption{Request latency breakdown under YCSB.}
    \label{latency-5}
  \end{minipage}
\end{figure}

As shown in Figure \ref{accesses-5}, Address Mapping requires an average of approximately 96 memory accesses per request under the balanced YCSB workload ($z=0.6$, $P=0.1$), about twice that of Transparent Expansion. To further analyze this performance difference, average request latency under the same workload is broken down into data distribution management, index lookup, and tuple access. Figure \ref{latency-5} shows that the index lookup overhead of Address Mapping is 56.62\% higher than that of Transparent Expansion and accounts for 91.92\% of the total request latency of Address Mapping. These results show that, despite explicit control over data distribution, the additional address resolution makes index lookup the primary source of latency. Consequently, Address Mapping achieves the lowest throughput among the five designs across all evaluated workloads.

Cached Mappings retains frequently accessed lookup-table entries in DDR memory to reduce the CXL memory accesses incurred by address resolution, achieving $1.04$--$1.16\times$ the throughput of Address Mapping. Under the same balanced workload, it reduces CXL memory accesses by 84.80\% relative to Address Mapping, but reduces index lookup overhead by only 16.36\%. This is because caching changes only the memory tier used for some address resolutions and does not eliminate address resolution itself. The average total number of memory accesses per request therefore remains approximately twice that of Transparent Expansion, while index lookup overhead remains 30.99\% higher. As a result, across all evaluated workloads, its throughput reaches only 81.31\%--87.21\% of that of Transparent Expansion. These results show that Cached Mappings mitigates address resolution overhead, but the additional accesses introduced by address resolution remain the main factor limiting its overall performance.

To address this limitation, Direct Addressing stores actual object addresses in the index and accesses objects directly through these addresses, eliminating address resolution from the data access path. Under the balanced workload, Direct Addressing reduces the average number of memory accesses per request by approximately 50\% relative to the two mapping-based designs, with corresponding reductions in index lookup overhead of 47.82\%--56.36\%. Direct Addressing and Transparent Expansion incur approximately the same average total number of memory accesses per request. By explicitly adjusting the distribution of index objects and tuples, Direct Addressing reduces CXL memory accesses by 59.06\% and index lookup overhead by 31.65\% relative to Transparent Expansion.

This simpler access path comes with additional migration maintenance overhead. Direct Addressing requires synchronized backpointer updates after an index object is migrated. Under the same workload, its average migration overhead is 34.81~ns per request, $1.59$--$2.09\times$ that of the two mapping-based designs. The relatively small latency gap between DDR and CXL limits the benefit obtained from each access after migration. Admission conditions derived from migration overhead and benefit per access therefore discourage migration of objects expected to receive few subsequent accesses. Consequently, migrations occur infrequently, and average data distribution management overhead remains below 100~ns per request for all four designs with explicit distribution management. Direct Addressing incurs 80.42~ns of data distribution management overhead per request. Although its migration overhead is relatively high, migration accounts for only 1.02\% of total request latency.

Therefore, under the same workload, Direct Addressing reduces total request latency by 42.42\%--50.93\% relative to the two mapping-based designs and by 27.19\% relative to Transparent Expansion. Across all evaluated workloads, it achieves $1.58$--$2.12\times$ the throughput of the two mapping-based designs and $1.35$--$1.64\times$ that of Transparent Expansion, delivering the highest throughput among the four single-index designs. These results show that, when the DDR--CXL access latency gap is small and benefit-aware admission can reduce unnecessary data migration, simplifying the frequently executed access path at the cost of higher migration maintenance overhead is a more favorable trade-off in indexing design.

The four preceding designs manage tuples distributed across DDR and CXL memory through a single index, whereas Dual Index maintains an index in each memory tier. Across all evaluated workloads, Dual Index achieves higher throughput than Transparent Expansion, Address Mapping, and Cached Mappings, reaching $1.21$--$2.07\times$ their throughput. Its performance relative to Direct Addressing improves as the Zipfian parameter $z$ increases. Direct Addressing achieves $1.08$--$1.19\times$ and $1.02$--$1.17\times$ the throughput of Dual Index under workloads with $z=0.3$ and $z=0.6$, respectively. Dual Index achieves $1.02$--$1.08\times$ the throughput of Direct Addressing in four workloads with $z=0.9$ and comparable throughput in the other workload.

This change in performance primarily arises from the shorter data access path that Dual Index provides for hot requests. The DDR index manages hot tuple copies, and a successful lookup in this index avoids a subsequent lookup in the complete index in CXL memory. Dual Index reduces the average number of memory accesses per request by 20.33\% relative to Direct Addressing under the balanced YCSB workload with $z=0.6$. As $z$ increases, requests become more concentrated on a small number of hot tuples, allowing more requests to use the shorter access path through the DDR index. Dual Index can therefore achieve a greater performance advantage under workloads with more concentrated accesses.

\subsection{Impact of CXL Memory Latency}
\label{sec:cxl-latency}

\begin{figure}[t]
  \centering
  \begin{minipage}{0.48\textwidth}
    \centering
    \includegraphics[width=\linewidth]{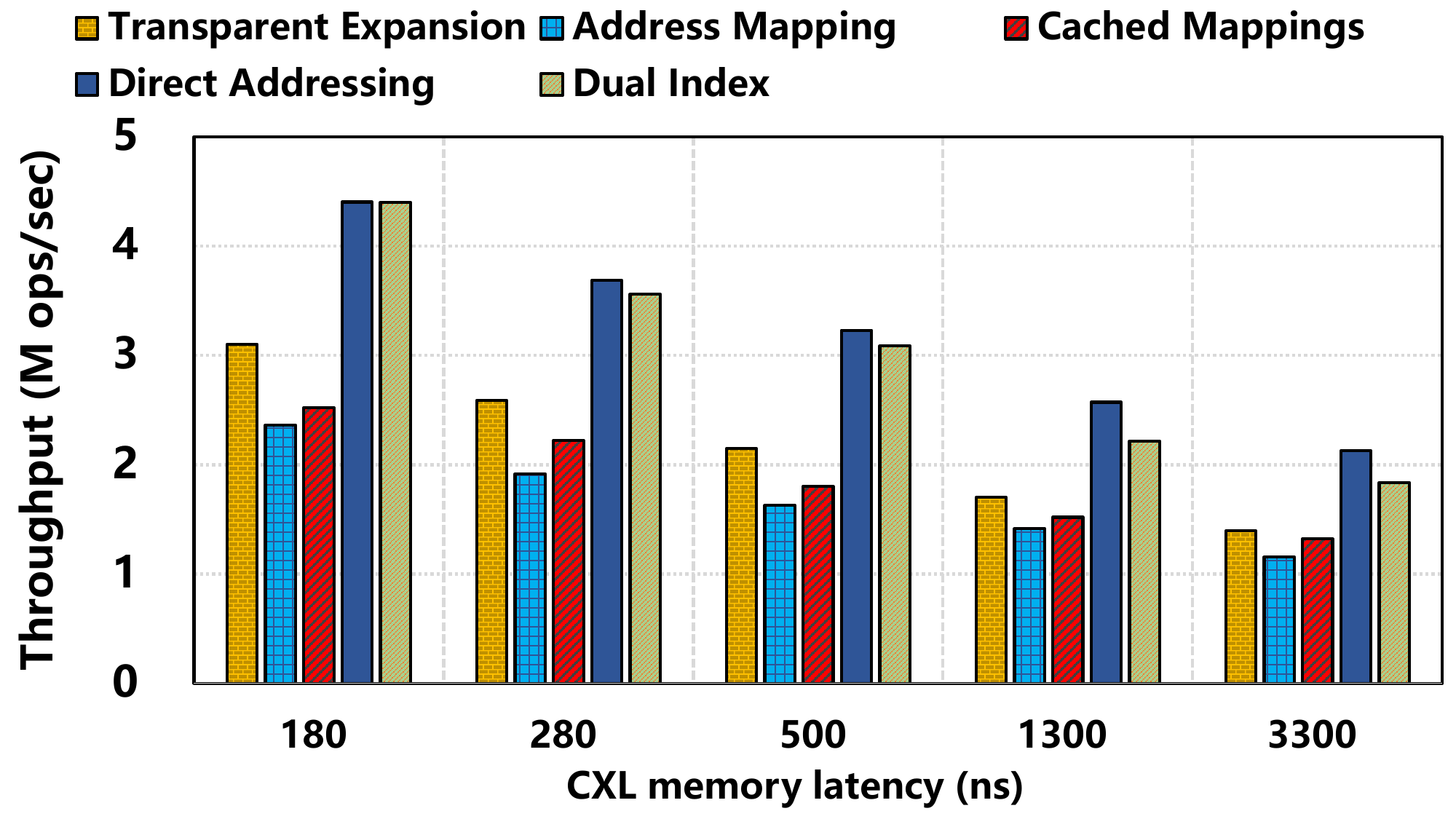}
    \caption{Robustness to CXL memory latency under YCSB.}
    \label{cxl-sim}
  \end{minipage}%
  \hfill
  \begin{minipage}{0.48\textwidth}
    \centering
    \includegraphics[width=\linewidth]{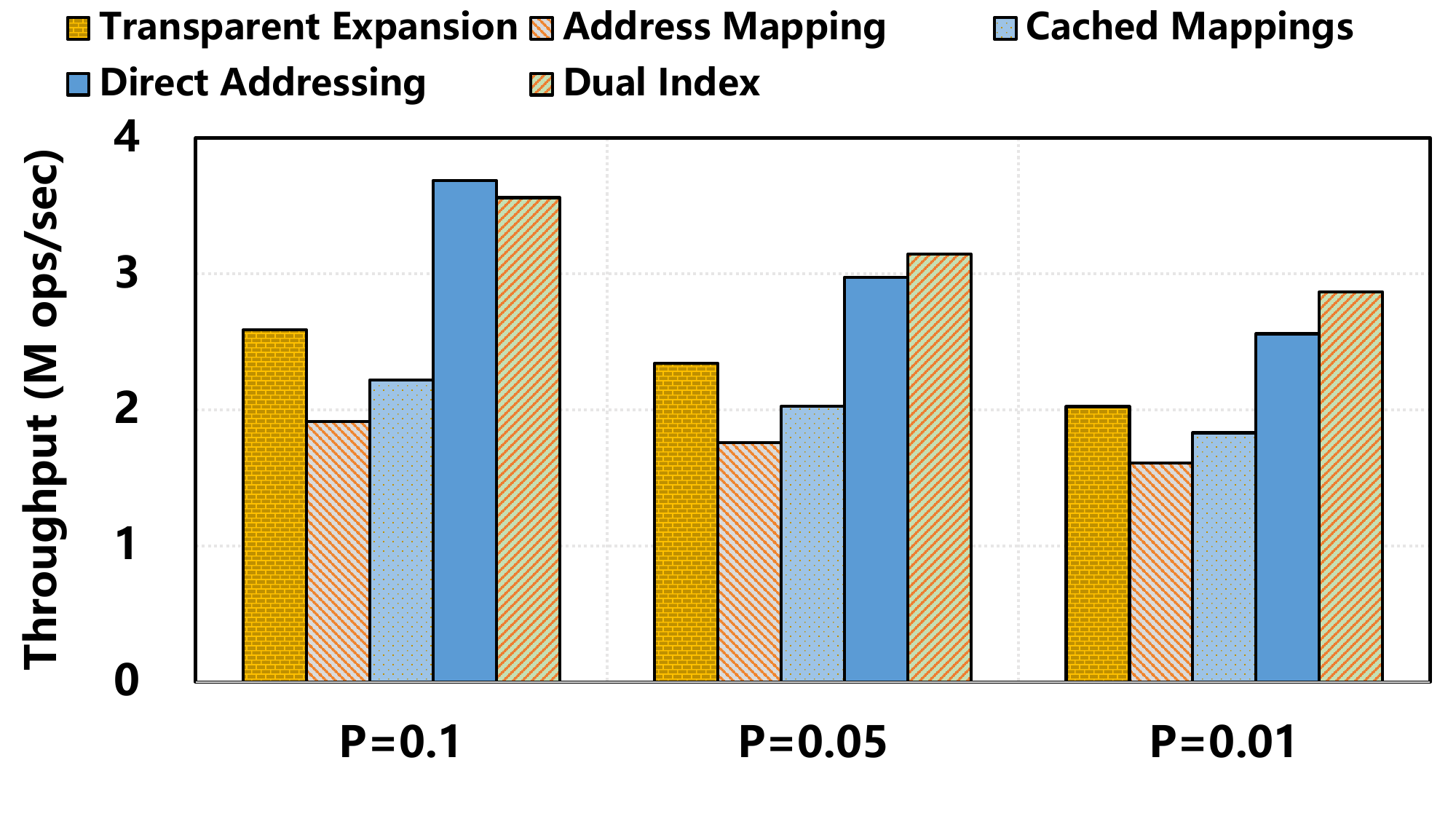}
    \caption{DDR memory efficiency under YCSB.}
    \label{cap-5}
  \end{minipage}
\end{figure}

This experiment evaluates the impact of CXL memory access latency on the performance of the five indexing designs. A balanced YCSB workload with $z=0.6$ and $P=0.1$ is used, with five CXL memory access latency settings: 180~ns, 280~ns, 500~ns, 1.3~$\mu$s, and 3.3~$\mu$s. These settings cover the latency characteristics of various systems, including systems based on CXL switches and CXL~3.0 \cite{switch1,switch2,switch3}. As shown in Figure \ref{cxl-sim}, throughput decreases for all five designs as CXL memory access latency increases, while their relative ranking remains unchanged. Direct Addressing achieves the highest throughput under all latency settings, whereas Address Mapping and Cached Mappings consistently achieve lower throughput than the other three designs. Both mapping-based designs require lookup-table queries to resolve actual object addresses. The additional memory accesses introduced by address resolution therefore limit their overall performance across the evaluated CXL latencies.

Increasing CXL memory access latency from 180~ns to 3.3~$\mu$s reduces the throughput of Transparent Expansion and Dual Index by 55.04\% and 58.23\%, respectively, the largest reductions among the five designs. Data distribution in Transparent Expansion is managed by the operating system, so index objects and tuples resident in DDR memory do not necessarily correspond to current access hotspots. Its access paths therefore contain relatively more CXL memory accesses. When a DDR index lookup in Dual Index is unsuccessful, a subsequent lookup in the complete index in CXL memory is required, also generating relatively more CXL memory accesses. By contrast, Direct Addressing can adjust the distribution of index objects and tuples separately. Even when the target tuple resides in CXL memory, some index objects along the lookup path can reside in DDR memory. Consequently, increasing CXL memory access latency incurs higher latency overhead for the CXL accesses in Transparent Expansion and Dual Index.

\subsection{Impact of DDR Memory Capacity}
\label{sec:ddr-capacity}

This experiment evaluates the impact of DDR memory capacity on the performance of the five indexing designs. A balanced YCSB workload with $z=0.6$ is used, with three DDR capacity configurations: $P=0.1$, $P=0.05$, and $P=0.01$. As shown in Figure \ref{cap-5}, reducing $P$ from 0.1 to 0.01 decreases throughput for all five designs. Address Mapping and Cached Mappings exhibit the smallest reductions, at 16.06\% and 17.46\%, respectively, whereas Direct Addressing exhibits the largest reduction, at 30.56\%.

The throughput of the two mapping-based designs is primarily limited by address resolution overhead, making the impact of reduced DDR memory capacity relatively small. Their throughput remains lower than that of the other three designs under all three capacity configurations. By contrast, Direct Addressing exhibits the largest throughput reduction, primarily because it maintains backpointers to record the locations of all pointers referencing an index object. These pointers must be updated after index object migration to ensure that subsequent requests can still locate the object correctly. Backpointers increase the memory footprint of index objects. Consequently, Direct Addressing can retain fewer hot index objects and tuples than the other designs under the same DDR capacity. As DDR capacity decreases, this additional space accounts for a larger fraction of the available capacity, leading to a more pronounced throughput reduction.

These differences in capacity sensitivity change both the relative performance of the five designs and which design achieves the highest throughput as DDR capacity varies, further reflecting their different DDR memory efficiencies. Direct Addressing achieves the highest throughput with $P=0.1$, reaching $1.04$--$1.92\times$ that of the other four designs. When $P$ decreases to 0.05 and 0.01, Dual Index achieves the highest throughput, reaching $1.06$--$1.79\times$ and $1.12$--$1.78\times$ that of the other four designs, respectively.

\subsection{Range Query Performance}
\label{sec:range-query}

\begin{figure*}[t]
\centering
\includegraphics[width=0.5\textwidth]{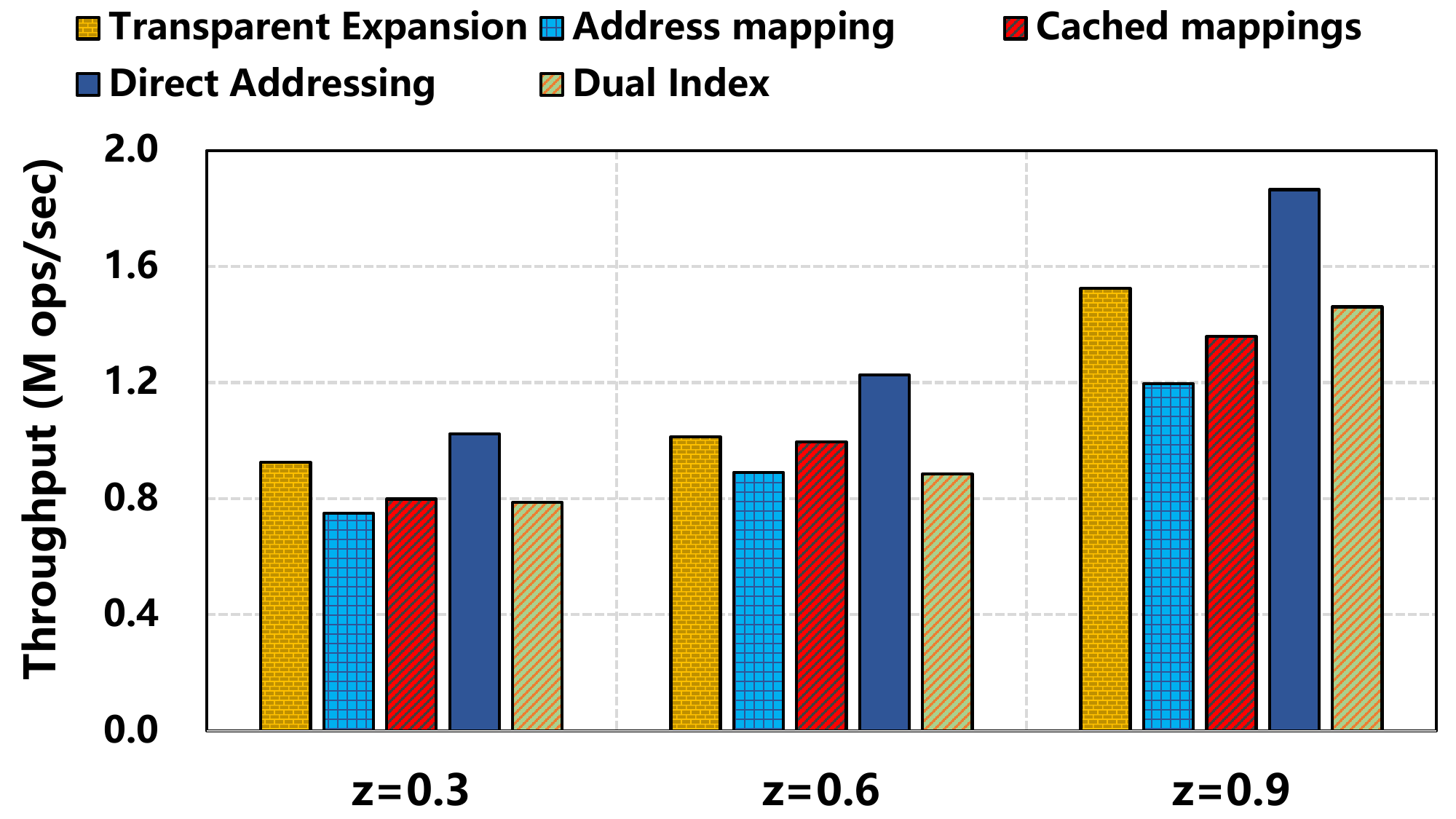}
\caption{Range query throughput under YCSB.}
\label{scan-5}
\end{figure*}

This experiment evaluates range query performance for the five indexing designs. A YCSB range query workload is used, with each request reading 10 consecutive tuples. The starting key follows a Zipfian distribution with $z=0.3$, $z=0.6$, or $z=0.9$. All other configurations are identical to those of the overall performance experiment. As shown in Figure \ref{scan-5}, the relative ranking of the four single-index designs under all three workload configurations is largely consistent with the overall performance results, with Direct Addressing achieving the highest throughput among them. However, the relative performance of Dual Index declines substantially for range queries.

These results primarily arise from the different execution characteristics of range queries and single-record operations, which consequently impose different requirements on indexing designs. For a single-index design, a range query must traverse subsequent index entries and access the corresponding tuples in sequence after locating the starting entry. The two mapping-based designs still require lookup-table queries to resolve actual object addresses during this process, resulting in lower throughput than Transparent Expansion, which accesses objects directly through their actual addresses. Direct Addressing also accesses objects through actual addresses and additionally allows the database to explicitly adjust the distribution of index objects and tuples. It therefore achieves $1.11$--$1.56\times$ the throughput of the other four designs. Dual Index, however, must traverse the complete index in CXL memory and query the DDR index to obtain the latest versions of hot tuples. It therefore cannot use the DDR index to shorten range query traversal and incurs additional data synchronization overhead.

\subsection{Discussion of Indexing Design Trade-offs}
\label{sec:indexing-discussion}

The preceding experiments show that, when the DDR--CXL access latency gap is small and benefit-aware admission can reduce unnecessary data migration, Direct Addressing offers a more favorable indexing trade-off by simplifying the frequently executed data access path at the cost of higher migration maintenance overhead. Both mapping-based designs require address resolution during object accesses. Cached Mappings changes only the memory tier used for some address resolutions without eliminating the additional accesses. Consequently, the two mapping-based designs that adopt the indirect addressing approach of conventional buffer management achieve lower performance than Transparent Expansion throughout the evaluated experiments.

In terms of runtime performance, Direct Addressing achieves the best performance under most workloads and system configurations, whereas Dual Index achieves higher performance when accesses are highly concentrated or DDR memory capacity is limited. Dual Index uses the DDR index to provide a shorter data access path for hot requests, with benefits increasing as the proportion of hot requests grows. However, the DDR index manages only the latest copies of hot tuples, so range queries still require traversal of the complete index in CXL memory. This advantage therefore does not extend to range queries. Direct Addressing incurs additional space overhead for backpointers, limiting its relative performance when DDR memory capacity is small. The relative ranking of the five designs remains unchanged across the evaluated CXL memory access latencies.

In terms of system integration complexity, Transparent Expansion requires no changes to the internal database implementation, but cannot explicitly control data distribution according to access characteristics and exhibits relatively low overall performance. Although Direct Addressing provides the runtime performance advantages described above, it requires an additional set of backpointers for each index object to record the locations of all pointers referencing that object. These backpointers must be updated synchronously during index object migration and all operations that may change the index structure. Integrating Direct Addressing therefore requires extensive changes to the existing index structure and its maintenance logic. Dual Index requires only the common modifications needed for explicit data distribution management. It achieves performance comparable to Direct Addressing under most single-record workloads and higher performance under some workload and capacity configurations. Dual Index is therefore the best choice when both system integration complexity and runtime performance are considered.

\begin{figure*}[t]
\centering
\includegraphics[width=\textwidth]{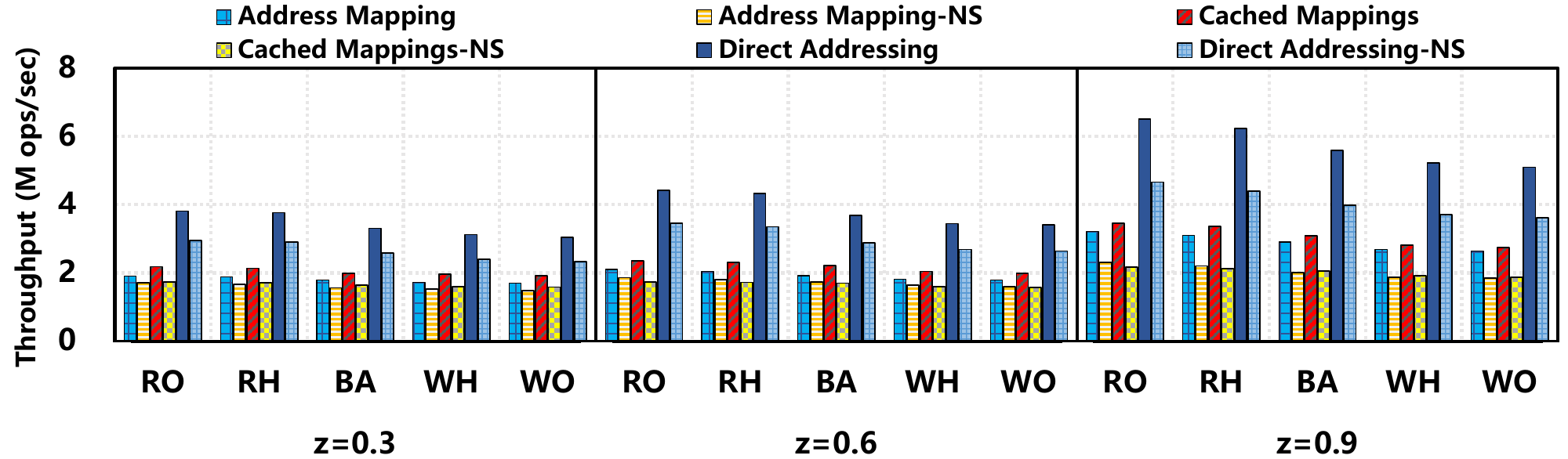}
\caption{Overall throughput under YCSB with separate and non-separated management.}
\label{overall-6}
\end{figure*}

\subsection{Impact of Separate Management of Index Objects and Tuples}
\label{sec:separate-management}

This ablation study evaluates the effectiveness of managing index objects and tuples separately. All comparisons use the workload configurations of the corresponding preceding experiments. The performance of Address Mapping, Cached Mappings, and Direct Addressing is compared under separate management and management of complete index nodes (non-separated, NS). Under non-separated management, the database still explicitly determines and adjusts data distribution, but the index object and tuple of each node always reside in the same memory tier. The implementations with non-separated management are denoted Address Mapping-NS, Cached Mappings-NS, and Direct Addressing-NS, respectively.

\begin{figure}[t]
  \centering
  \begin{minipage}{0.48\textwidth}
    \centering
    \includegraphics[width=\linewidth]{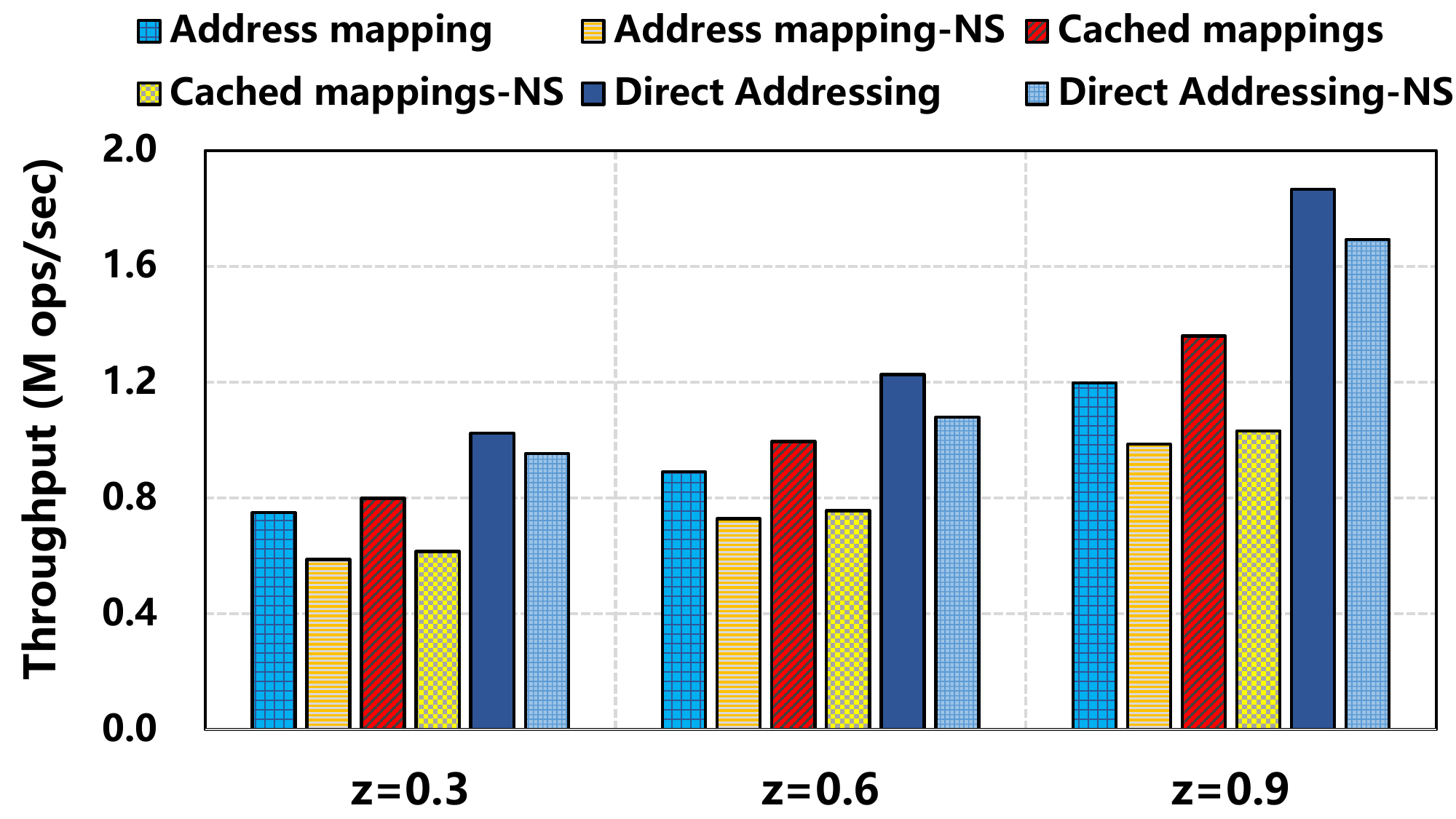}
    \caption{Range query throughput under YCSB.}
    \label{scan-6}
  \end{minipage}%
  \hfill
  \begin{minipage}{0.48\textwidth}
    \centering
    \includegraphics[width=\linewidth]{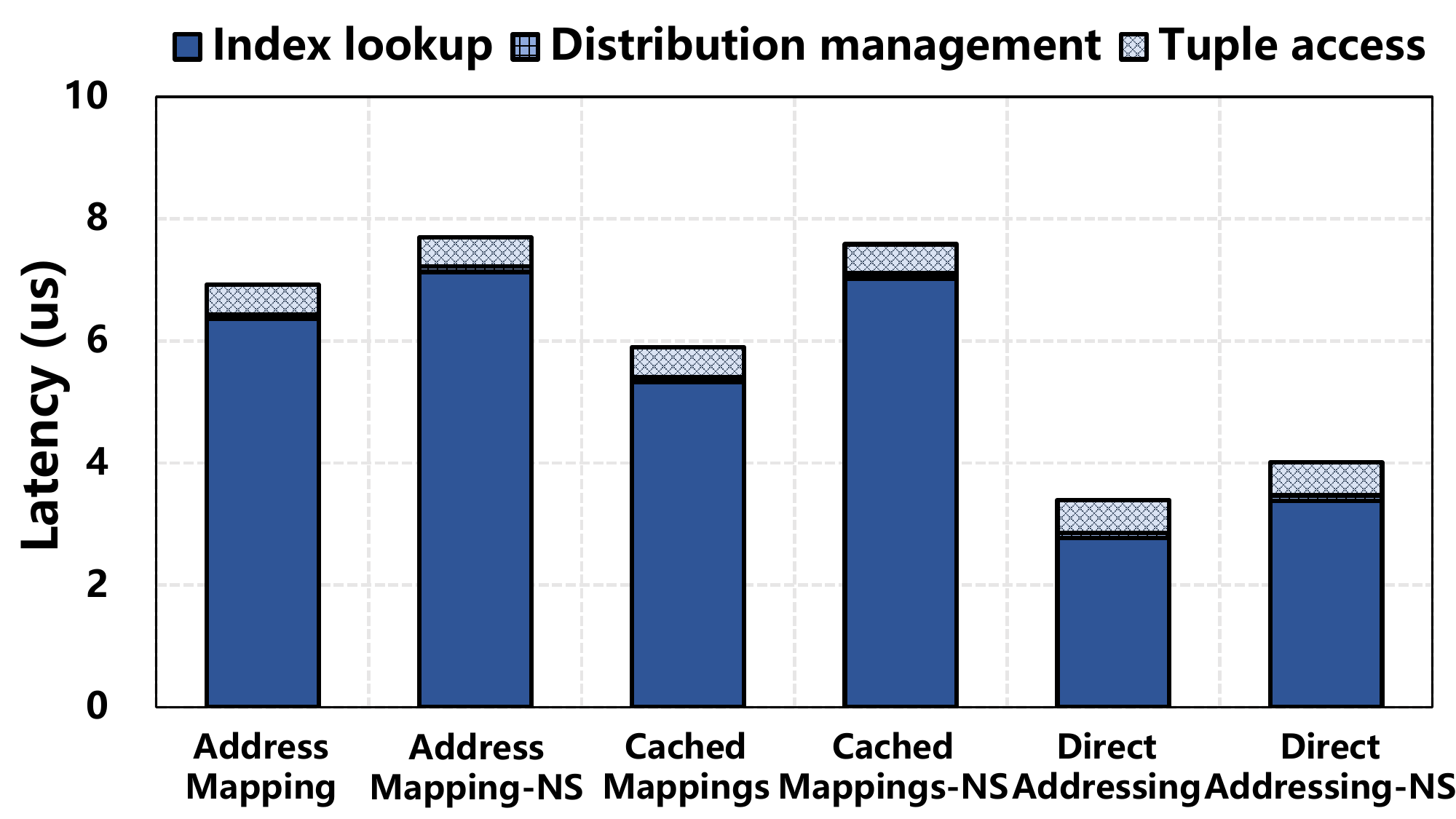}
    \caption{Request latency breakdown under YCSB.}
    \label{latency-6}
  \end{minipage}
\end{figure}

As shown in Figure \ref{overall-6}, all three designs achieve higher throughput with separate management across the evaluated read/write ratios and Zipfian parameters. Address Mapping, Cached Mappings, and Direct Addressing achieve $1.11$--$1.44\times$, $1.21$--$1.59\times$, and $1.28$--$1.42\times$ the throughput of their corresponding implementations with non-separated management, respectively. Separate management also improves range query performance. As shown in Figure \ref{scan-6}, the corresponding throughput ratios are $1.21$--$1.28\times$, $1.30$--$1.32\times$, and $1.07$--$1.14\times$, respectively. Figure \ref{latency-6} shows that, under the balanced YCSB workload ($z=0.6$, $P=0.1$), separate management also reduces total request latency by 10.04\%, 22.23\%, and 15.36\%, respectively. The corresponding reductions in index lookup overhead are 10.77\%, 24.25\%, and 17.81\%.

Although tracking index objects and tuples separately increases hotness maintenance overhead, average data distribution management overhead remains below 100~ns per request for all three designs. Hierarchical hotness tracking maintains fine-grained hotness statistics for index objects only in active logical partitions and uses index object hotness to select tuples for tracking, thereby limiting the scope of fine-grained hotness tracking. The index object residency policy further keeps index objects in DDR memory when their hotness reaches the saturation bound but their corresponding tuples have not yet met the admission condition. The hotness of these objects is more likely to accumulate from accesses along multiple lookup paths and is less sensitive to changes in accesses to an individual tuple. Keeping these objects resident can eliminate their hotness tracking and migration overheads.

\begin{figure}[htbp]
  \centering
  \begin{minipage}{0.48\textwidth}
    \centering
    \includegraphics[width=\linewidth]{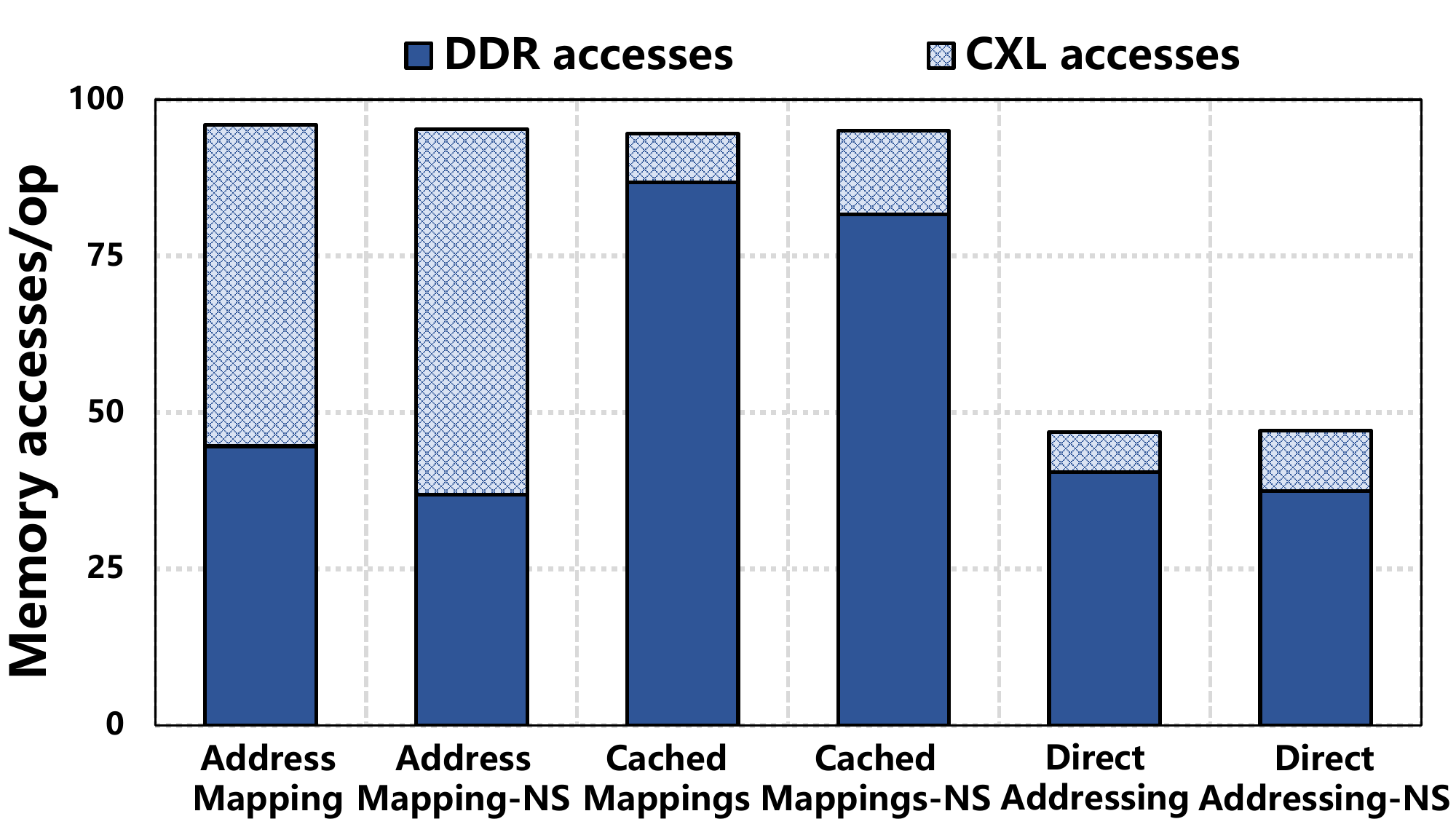}
    \caption{Memory accesses per request under YCSB.}
    \label{accesses-6}
  \end{minipage}%
  \hfill
  \begin{minipage}{0.48\textwidth}
    \centering
    \includegraphics[width=\linewidth]{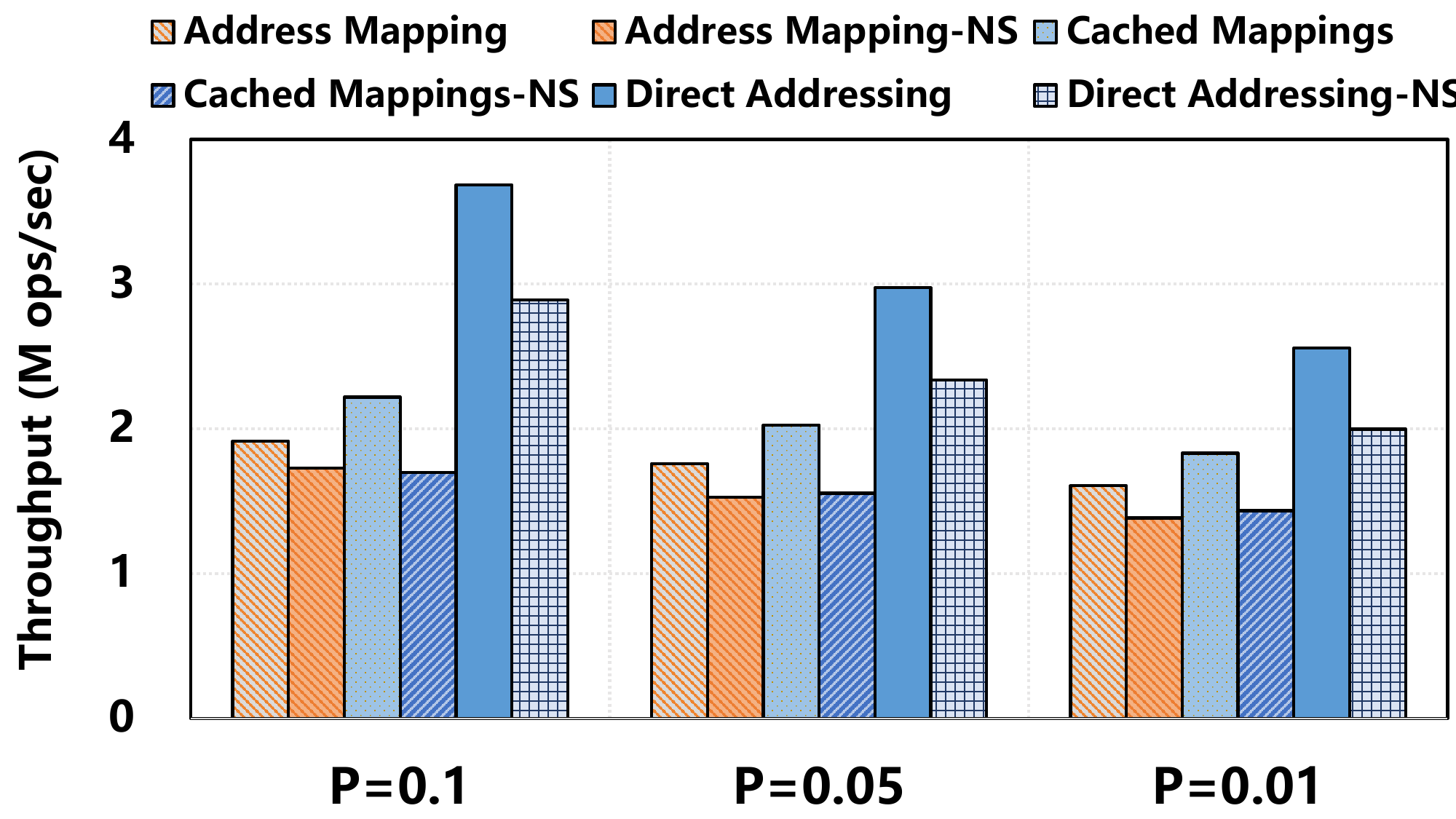}
    \caption{DDR memory efficiency under YCSB.}
    \label{cap-6}
  \end{minipage}
\end{figure}

Separate management also improves DDR memory efficiency. As shown in Figure \ref{accesses-6}, it allows more memory accesses to be served by DDR while the average total number of memory accesses per request remains approximately unchanged. CXL memory accesses decrease by 12.14\%, 41.78\%, and 33.63\% for Address Mapping, Cached Mappings, and Direct Addressing, respectively, which also explains the lower index lookup overhead. Figure \ref{cap-6} further shows that, across DDR capacity configurations of $P=0.1$, $P=0.05$, and $P=0.01$, Address Mapping, Cached Mappings, and Direct Addressing achieve $1.11$--$1.16\times$, $1.28$--$1.31\times$, and $1.27$--$1.28\times$ the throughput of their corresponding implementations with non-separated management, respectively. 

This difference primarily arises because, although both management approaches adjust data distribution according to expected benefit, non-separated management requires the index object and tuple of each node to reside in the same memory tier. When their access characteristics differ, the component with higher expected benefit may cause the entire node to be placed in DDR memory, while the other component consequently reduces DDR memory efficiency. In contrast, separate management evaluates admission benefits independently for index objects and tuples and considers their memory footprints during eviction, allowing limited DDR capacity to be used preferentially to retain objects with higher expected benefit.

\section{Conclusion}
In this paper, we propose and implement five indexing designs for hybrid DDR--CXL databases and systematically analyze their differences and trade-offs in request execution, data distribution adjustment, DDR memory efficiency, and system integration complexity. We also design an adaptive data distribution optimization method that manages index objects and tuples separately in single-index designs. The method reduces metadata management overhead through hierarchical hotness tracking and an index object residency policy, and determines and adjusts data distribution according to the respective expected benefits and memory footprints of the two object types. Experimental results show that the single-index design with direct addressing achieves the best performance under most workloads, and that separate management of index objects and tuples improves DDR memory efficiency and system performance. This work provides a basis for selecting indexing designs and optimizing data distribution in hybrid DDR--CXL databases under different workloads and memory configurations.

\bibliographystyle{ACM-Reference-Format}
\bibliography{sample-base}

@article{hotspot_11,
author = {Jin, Hai and Li, Zhiwei and Liu, Haikun and Liao, Xiaofei and Zhang, Yu},
title = {Hotspot-Aware Hybrid Memory Management for In-Memory Key-Value Stores},
year = {2020},
issue_date = {April 2020},
xpublisher = {IEEE Press},
volume = {31},
number = {4},
issn = {1045-9219},
xurl = {https://doi.org/10.1109/TPDS.2019.2945315},
xdoi = {10.1109/TPDS.2019.2945315},
journal = {IEEE Transactions on Parallel and Distributed Systems},
month = apr,
pages = {779–792},
numpages = {14}
}

@inproceedings{ycsb,
  title={Benchmarking cloud serving systems with YCSB},
  author={Cooper, Brian F and Silberstein, Adam and Tam, Erwin and Ramakrishnan, Raghu and Sears, Russell},
  booktitle={Proceedings of the 1st ACM symposium on Cloud computing},
  pages={143--154},
  year={2010},
xpublisher = {Association for Computing Machinery},
xaddress = {New York, NY, USA},
}

@article{anti,
author = {DeBrabant, Justin and Pavlo, Andrew and Tu, Stephen and Stonebraker, Michael and Zdonik, Stan},
title = {Anti-caching: a new approach to database management system architecture},
year = {2013},
issue_date = {September 2013},
xpublisher = {VLDB Endowment},
volume = {6},
number = {14},
issn = {2150-8097},
xurl = {https://doi.org/10.14778/2556549.2556575},
xdoi = {10.14778/2556549.2556575},
journal = {Proceedings of the VLDB Endowment},
month = sep,
pages = {1942–1953},
numpages = {12}
}

@inproceedings {nomad,
author = {Lingfeng Xiang and Zhen Lin and Weishu Deng and Hui Lu and Jia Rao and Yifan Yuan and Ren Wang},
title = {Nomad: {Non-Exclusive} Memory Tiering via Transactional Page Migration},
booktitle = {18th USENIX Symposium on Operating Systems Design and Implementation},
year = {2024},
isbn = {978-1-939133-40-3},
xaddress = {Santa Clara, CA},
pages = {19--35},
xurl = {https://www.usenix.org/conference/osdi24/presentation/xiang},
xpublisher = {USENIX Association},
month = jul
}

@inproceedings{hybridtier,
author = {Song, Kevin and Yang, Jiacheng and Wang, Zixuan and Zhao, Jishen and Liu, Sihang and Pekhimenko, Gennady},
title = {HybridTier: an Adaptive and Lightweight CXL-Memory Tiering System},
year = {2025},
isbn = {9798400710803},
xpublisher = {Association for Computing Machinery},
xaddress = {New York, NY, USA},
xurl = {https://doi.org/10.1145/3676642.3736119},
xdoi = {10.1145/3676642.3736119},
booktitle = {Proceedings of the 30th ACM International Conference on Architectural Support for Programming Languages and Operating Systems, Volume 3},
pages = {112–128},
numpages = {17},
location = {Rotterdam, Netherlands},
xseries = {ASPLOS '25}
}

@article{Siberia,
author = {Eldawy, Ahmed and Levandoski, Justin and Larson, Per-Ake},
title = {Trekking through Siberia: managing cold data in a memory-optimized database},
year = {2014},
issue_date = {July 2014},
xpublisher = {VLDB Endowment},
volume = {7},
number = {11},
issn = {2150-8097},
xurl = {https://doi.org/10.14778/2732967.2732968},
xdoi = {10.14778/2732967.2732968},
journal = {Proceedings of the VLDB Endowment},
month = jul,
pages = {931–942},
numpages = {12}
}

@inproceedings{Siberia2,
author = {Diaconu, Cristian and Freedman, Craig and Ismert, Erik and Larson, Per-Ake and Mittal, Pravin and Stonecipher, Ryan and Verma, Nitin and Zwilling, Mike},
title = {Hekaton: SQL server's memory-optimized OLTP engine},
year = {2013},
isbn = {9781450320375},
xpublisher = {Association for Computing Machinery},
xaddress = {New York, NY, USA},
xurl = {https://doi.org/10.1145/2463676.2463710},
xdoi = {10.1145/2463676.2463710},
booktitle = {Proceedings of the 2013 ACM SIGMOD International Conference on Management of Data},
pages = {1243–1254},
numpages = {12},
location = {New York, New York, USA},
xseries = {SIGMOD '13}
}

@article{Bf-Tree,
author = {Hao, Xiangpeng and Chandramouli, Badrish},
title = {Bf-Tree: A Modern Read-Write-Optimized Concurrent Larger-Than-Memory Range Index},
year = {2024},
issue_date = {July 2024},
xpublisher = {VLDB Endowment},
volume = {17},
number = {11},
issn = {2150-8097},
xurl = {https://doi.org/10.14778/3681954.3682012},
xdoi = {10.14778/3681954.3682012},
journal = {Proceedings of the VLDB Endowment},
month = jul,
pages = {3442–3455},
numpages = {14}
}

@article{Tiered-Indexing,
author = {Zhou, Xinjing and Hao, Xiangpeng and Yu, Xiangyao and Stonebraker, Michael},
title = {Tiered-Indexing: Optimizing Access Methods for Skew},
year = {2025},
issue_date = {Jun 2025},
xpublisher = {Springer-Verlag},
xaddress = {Berlin, Heidelberg},
volume = {34},
number = {4},
issn = {1066-8888},
xurl = {https://doi.org/10.1007/s00778-025-00928-6},
xdoi = {10.1007/s00778-025-00928-6},
journal = {The VLDB Journal},
month = may,
numpages = {24}
}

@inproceedings{FOEDUS,
author = {Kimura, Hideaki},
title = {FOEDUS: OLTP Engine for a Thousand Cores and NVRAM},
year = {2015},
isbn = {9781450327589},
xpublisher = {Association for Computing Machinery},
xaddress = {New York, NY, USA},
xurl = {https://doi.org/10.1145/2723372.2746480},
xdoi = {10.1145/2723372.2746480},
booktitle = {Proceedings of the 2015 ACM SIGMOD International Conference on Management of Data},
pages = {691–706},
numpages = {16},
location = {Melbourne, Victoria, Australia},
xseries = {SIGMOD '15}
}

@inproceedings{FPTree,
author = {Oukid, Ismail and Lasperas, Johan and Nica, Anisoara and Willhalm, Thomas and Lehner, Wolfgang},
title = {FPTree: A Hybrid SCM-DRAM Persistent and Concurrent B-Tree for Storage Class Memory},
year = {2016},
isbn = {9781450335317},
xpublisher = {Association for Computing Machinery},
xaddress = {New York, NY, USA},
xurl = {https://doi.org/10.1145/2882903.2915251},
xdoi = {10.1145/2882903.2915251},
booktitle = {Proceedings of the 2016 International Conference on Management of Data},
pages = {371–386},
numpages = {16},
location = {San Francisco, California, USA},
xseries = {SIGMOD '16}
}

@inproceedings{HiKV,
author = {Xia, Fei and Jiang, Dejun and Xiong, Jin and Sun, Ninghui},
title = {HiKV: a hybrid index key-value store for DRAM-NVM memory systems},
year = {2017},
isbn = {9781931971386},
xpublisher = {USENIX Association},
xaddress = {USA},
booktitle = {Proceedings of the 2017 USENIX Conference on Usenix Annual Technical Conference},
pages = {349–362},
numpages = {14},
location = {Santa Clara, CA, USA},
xseries = {USENIX ATC '17}
}

@inproceedings{two,
  title={Two is better than one: The case for 2-tree for skewed data sets},
  author={Zhou, Xinjing and Yu, Xiangyao and Graefe, Goetz and Stonebraker, Michael},
  booktitle={13th Annual Conference on Innovative Data Systems Research},
  year={2023}
}

@inproceedings{fifo,
author = {Yang, Juncheng and Zhang, Yazhuo and Qiu, Ziyue and Yue, Yao and Vinayak, Rashmi},
title = {FIFO queues are all you need for cache eviction},
year = {2023},
isbn = {9798400702297},
xpublisher = {Association for Computing Machinery},
xaddress = {New York, NY, USA},
xurl = {https://doi.org/10.1145/3600006.3613147},
xdoi = {10.1145/3600006.3613147},
booktitle = {Proceedings of the 29th Symposium on Operating Systems Principles},
pages = {130–149},
numpages = {20},
location = {Koblenz, Germany},
xseries = {SOSP '23}
}

@article{TinyLFU,
author = {Einziger, Gil and Friedman, Roy and Manes, Ben},
title = {TinyLFU: A Highly Efficient Cache Admission Policy},
year = {2017},
issue_date = {November 2017},
xpublisher = {Association for Computing Machinery},
xaddress = {New York, NY, USA},
volume = {13},
number = {4},
issn = {1553-3077},
xurl = {https://doi.org/10.1145/3149371},
xdoi = {10.1145/3149371},
journal = {ACM Transactions on Storage},
month = nov,
articleno = {35},
numpages = {31}
}

@article{DPTree,
author = {Zhou, Xinjing and Shou, Lidan and Chen, Ke and Hu, Wei and Chen, Gang},
title = {DPTree: differential indexing for persistent memory},
year = {2019},
issue_date = {December 2019},
xpublisher = {VLDB Endowment},
volume = {13},
number = {4},
issn = {2150-8097},
xurl = {https://doi.org/10.14778/3372716.3372717},
xdoi = {10.14778/3372716.3372717},
journal = {Proceedings of the VLDB Endowment},
month = dec,
pages = {421–434},
numpages = {14}
}

@article{BonsaiKV,
author = {Cai, Miao and Shen, Junru and Yuan, Yifan and Qu, Zhihao and Ye, Baoliu},
title = {BonsaiKV: Towards Fast, Scalable, and Persistent Key-Value Stores with Tiered, Heterogeneous Memory System},
year = {2023},
issue_date = {December 2023},
xpublisher = {VLDB Endowment},
volume = {17},
number = {4},
issn = {2150-8097},
xurl = {https://doi.org/10.14778/3636218.3636228},
xdoi = {10.14778/3636218.3636228},
journal = {Proceedings of the VLDB Endowment},
month = dec,
pages = {726–739},
numpages = {14}
}

@article{Nap,
author = {Wang, Qing and Lu, Youyou and Li, Junru and Xie, Minhui and Shu, Jiwu},
title = {Nap: Persistent Memory Indexes for NUMA Architectures},
year = {2022},
issue_date = {February 2022},
xpublisher = {Association for Computing Machinery},
xaddress = {New York, NY, USA},
volume = {18},
number = {1},
issn = {1553-3077},
xurl = {https://doi.org/10.1145/3507922},
xdoi = {10.1145/3507922},
journal = {ACM Transactions on Storage},
month = jan,
articleno = {2},
numpages = {35}
}

@article{DEX,
author = {Lu, Baotong and Huang, Kaisong and Liang, Chieh-Jan Mike and Wang, Tianzheng and Lo, Eric},
title = {DEX: Scalable Range Indexing on Disaggregated Memory},
year = {2024},
issue_date = {June 2024},
xpublisher = {VLDB Endowment},
volume = {17},
number = {10},
issn = {2150-8097},
xurl = {https://doi.org/10.14778/3675034.3675050},
xdoi = {10.14778/3675034.3675050},
journal = {Proceedings of the VLDB Endowment},
month = jun,
pages = {2603–2616},
numpages = {14}
}

@inproceedings{CHIME,
author = {Luo, Xuchuan and Shen, Jiacheng and Zuo, Pengfei and Wang, Xin and Lyu, Michael R. and Zhou, Yangfan},
title = {CHIME: A Cache-Efficient and High-Performance Hybrid Index on Disaggregated Memory},
year = {2024},
isbn = {9798400712517},
xpublisher = {Association for Computing Machinery},
xaddress = {New York, NY, USA},
xurl = {https://doi.org/10.1145/3694715.3695959},
xdoi = {10.1145/3694715.3695959},
booktitle = {Proceedings of the ACM SIGOPS 30th Symposium on Operating Systems Principles},
pages = {110–126},
numpages = {17},
location = {Austin, TX, USA},
xseries = {SOSP '24}
}

@inproceedings{ziegler2019rdmaindex,
author = {Ziegler, Tobias and Tumkur Vani, Sumukha and Binnig, Carsten and Fonseca, Rodrigo and Kraska, Tim},
title = {Designing Distributed Tree-based Index Structures for Fast RDMA-capable Networks},
year = {2019},
isbn = {9781450356435},
xpublisher = {Association for Computing Machinery},
xaddress = {New York, NY, USA},
xurl = {https://doi.org/10.1145/3299869.3300081},
xdoi = {10.1145/3299869.3300081},
booktitle = {Proceedings of the 2019 International Conference on Management of Data},
pages = {741–758},
numpages = {18},
location = {Amsterdam, Netherlands},
xseries = {SIGMOD '19}
}

@inproceedings{Deft,
author = {Wang, Jing and Wang, Qing and Zhang, Yuhao and Shu, Jiwu},
title = {Deft: A Scalable Tree Index for Disaggregated Memory},
year = {2025},
isbn = {9798400711961},
xpublisher = {Association for Computing Machinery},
xaddress = {New York, NY, USA},
xurl = {https://doi.org/10.1145/3689031.3696062},
xdoi = {10.1145/3689031.3696062},
booktitle = {Proceedings of the Twentieth European Conference on Computer Systems},
pages = {886–901},
numpages = {16},
location = {Rotterdam, Netherlands},
xseries = {EuroSys '25}
}

@inproceedings{Sherman,
author = {Wang, Qing and Lu, Youyou and Shu, Jiwu},
title = {Sherman: A Write-Optimized Distributed B+Tree Index on Disaggregated Memory},
year = {2022},
isbn = {9781450392495},
xpublisher = {Association for Computing Machinery},
xaddress = {New York, NY, USA},
xurl = {https://doi.org/10.1145/3514221.3517824},
xdoi = {10.1145/3514221.3517824},
booktitle = {Proceedings of the 2022 International Conference on Management of Data},
pages = {1033–1048},
numpages = {16},
location = {Philadelphia, PA, USA},
xseries = {SIGMOD '22}
}

@article{Hao,
author = {Hao, Xiangpeng and Zhou, Xinjing and Yu, Xiangyao and Stonebraker, Michael},
title = {Towards Buffer Management with Tiered Main Memory},
year = {2024},
issue_date = {February 2024},
xpublisher = {Association for Computing Machinery},
xaddress = {New York, NY, USA},
volume = {2},
number = {1},
xurl = {https://doi.org/10.1145/3639286},
xdoi = {10.1145/3639286},
journal = {Proceedings of the ACM on Management of Data},
month = mar,
articleno = {31},
numpages = {26}
}

@article{SIDLE,
author = {Zhao, Haoru and Dong, Mingkai and Wu, Fangnuo and Chen, Haibo},
title = {SIDLE: Tree-Structure Aware Indexes for CXL-Based Heterogeneous Memory},
year = {2026},
issue_date = {March 2026},
xpublisher = {VLDB Endowment},
volume = {19},
number = {7},
issn = {2150-8097},
xurl = {https://doi.org/10.14778/3801059.3801065},
xdoi = {10.14778/3801059.3801065},
journal = {Proceedings of the VLDB Endowment},
month = jul,
pages = {1499–1515},
numpages = {17}
}

@inproceedings {ROLEX,
author = {Li, Pengfei and Hua, Yu and Zuo, Pengfei and Chen, Zhangyu and Sheng, Jiajie},
title = {{ROLEX}: A Scalable {RDMA-oriented} Learned {Key-Value} Store for Disaggregated Memory Systems},
booktitle = {21st USENIX Conference on File and Storage Technologies},
year = {2023},
isbn = {978-1-939133-32-8},
xaddress = {Santa Clara, CA},
pages = {99--114},
xurl = {https://www.usenix.org/conference/fast23/presentation/li-pengfei},
xpublisher = {USENIX Association},
month = feb
}

@article{smart,
author = {Luo, Xuchuan and Zuo, Pengfei and Shen, Jiacheng and Gu, Jiazhen and Wang, Xin and Lyu, Michael and Zhou, Yangfan},
title = {A Memory-Disaggregated Radix Tree},
year = {2024},
issue_date = {August 2024},
xpublisher = {Association for Computing Machinery},
xaddress = {New York, NY, USA},
volume = {20},
number = {3},
issn = {1553-3077},
xurl = {https://doi.org/10.1145/3664289},
xdoi = {10.1145/3664289},
journal = {ACM Transactions on Storage},
month = jun,
articleno = {15},
numpages = {41}
}

@article{dLSM,
author = {Wang, Ruihong and Gao, Chuqing and Wang, Jianguo and Kadam, Prishita and Tamer{\"O}zsu, M. and Aref, Walid G.},
title = {Optimizing LSM-based indexes for disaggregated memory},
year = {2024},
issue_date = {Nov 2024},
xpublisher = {Springer-Verlag},
xaddress = {Berlin, Heidelberg},
volume = {33},
number = {6},
issn = {1066-8888},
xurl = {https://doi.org/10.1007/s00778-024-00863-y},
xdoi = {10.1007/s00778-024-00863-y},
journal = {The VLDB Journal},
month = jun,
pages = {1813–1836},
numpages = {24}
}

@ARTICLE{cxl,
  author={Sharma, Debendra Das},
  journal={IEEE Micro}, 
  title={Compute Express Link (CXL): Enabling Heterogeneous Data-Centric Computing With Heterogeneous Memory Hierarchy}, 
  year={2023},
  volume={43},
  number={2},
  pages={99-109},
  xdoi={10.1109/MM.2022.3228561}}

@INPROCEEDINGS{ddr1,
  author={Lee, Donghyuk and O'Connor, Mike and Chatterjee, Niladrish},
  booktitle={2018 IEEE International Symposium on High Performance Computer Architecture}, 
  title={Reducing Data Transfer Energy by Exploiting Similarity within a Data Transaction}, 
  year={2018},
  volume={},
  number={},
  pages={40-51},
  xdoi={10.1109/HPCA.2018.00014}}

@inproceedings{ddr2,
author = {Sun, Yan and Yuan, Yifan and Yu, Zeduo and Kuper, Reese and Song, Chihun and Huang, Jinghan and Ji, Houxiang and Agarwal, Siddharth and Lou, Jiaqi and Jeong, Ipoom and Wang, Ren and Ahn, Jung Ho and Xu, Tianyin and Kim, Nam Sung},
title = {Demystifying CXL Memory with Genuine CXL-Ready Systems and Devices},
year = {2023},
isbn = {9798400703294},
xpublisher = {Association for Computing Machinery},
xaddress = {New York, NY, USA},
xurl = {https://doi.org/10.1145/3613424.3614256},
xdoi = {10.1145/3613424.3614256},
booktitle = {Proceedings of the 56th Annual IEEE/ACM International Symposium on Microarchitecture},
pages = {105–121},
numpages = {17},
location = {Toronto, ON, Canada},
xseries = {MICRO '23}
}

@INPROCEEDINGS{im3,
  author={Kemper, Alfons and Neumann, Thomas},
  booktitle={2011 IEEE 27th International Conference on Data Engineering}, 
  title={HyPer: A hybrid OLTP\&OLAP main memory database system based on virtual memory snapshots}, 
  year={2011},
  volume={},
  number={},
  pages={195-206},
  xdoi={10.1109/ICDE.2011.5767867},
  ISSN={2375-026X},
  month={April}}

@inproceedings{ev1,
   title={Performance Characterization of CXL Memory and Its Use Cases},
   xurl={http://dx.doi.org/10.1109/IPDPS64566.2025.00097},
   xdoi={10.1109/ipdps64566.2025.00097},
   booktitle={2025 IEEE International Parallel and Distributed Processing Symposium},
   xpublisher={IEEE},
   author={Wang, Xi and Liu, Jie and Wu, Jianbo and Yang, Shuangyan and Ren, Jie and Shankar, Bhanu and Li, Dong},
   year={2025},
   month=jun, pages={1048–1061} }

@inproceedings{ev2,
author = {Tang, Yupeng and Zhou, Ping and Zhang, Wenhui and Hu, Henry and Yang, Qirui and Xiang, Hao and Liu, Tongping and Shan, Jiaxin and Huang, Ruoyun and Zhao, Cheng and Chen, Cheng and Zhang, Hui and Liu, Fei and Zhang, Shuai and Ding, Xiaoning and Chen, Jianjun},
title = {Exploring Performance and Cost Optimization with ASIC-Based CXL Memory},
year = {2024},
isbn = {9798400704376},
xpublisher = {Association for Computing Machinery},
xaddress = {New York, NY, USA},
xurl = {https://doi.org/10.1145/3627703.3650061},
xdoi = {10.1145/3627703.3650061},
booktitle = {Proceedings of the Nineteenth European Conference on Computer Systems},
pages = {818–833},
numpages = {16},
location = {Athens, Greece},
xseries = {EuroSys '24}
}

@article{zen,
author = {Liu, Gang and Chen, Leying and Chen, Shimin},
title = {Zen: a high-throughput log-free OLTP engine for non-volatile main memory},
year = {2021},
issue_date = {January 2021},
xpublisher = {VLDB Endowment},
volume = {14},
number = {5},
issn = {2150-8097},
xurl = {https://doi.org/10.14778/3446095.3446105},
xdoi = {10.14778/3446095.3446105},
journal = {Proceedings of the VLDB Endowment},
month = jan,
pages = {835–848},
numpages = {14}
}

@electronic{cache_shard_1,
  title         = "{RocksDB} Tuning Guide",
  year = {2023},
  url           = "https://github.com/facebook/rocksdb/wiki/RocksDB-Tuning-Guide"
 }

@electronic{cache_shard_2,
  title         = "Block Cache",
  year = {2023},
  url           = "https://github.com/facebook/rocksdb/wiki/Block-Cache"
 }

@electronic{decay_1,
  title         = "{Caffeine}: A high performance caching library for {Java}",
  year = {2026},
  url           = "https://github.com/ben-manes/caffeine"
 }

@electronic{mlc,
  title         = "Intel® Memory Latency Checker v3.12",
  year = {2025},
  url           = "https://www.intel.com/content/www/us/en/developer/articles/tool/intelr-memory-latency-checker.html"
 }

@misc{postgresql_buffer,
  title  = {{PostgreSQL} Buffer Manager},
  year   = {2026},
  url    = {https://github.com/postgres/postgres/blob/master/src/backend/storage/buffer/README}
}

@article{sqlite,
author = {Gaffney, Kevin P. and Prammer, Martin and Brasfield, Larry and Hipp, D. Richard and Kennedy, Dan and Patel, Jignesh M.},
title = {SQLite: past, present, and future},
year = {2022},
issue_date = {August 2022},
xpublisher = {VLDB Endowment},
volume = {15},
number = {12},
issn = {2150-8097},
xurl = {https://doi.org/10.14778/3554821.3554842},
xdoi = {10.14778/3554821.3554842},
journal = {Proceedings of the VLDB Endowment},
month = aug,
pages = {3535–3547},
numpages = {13}
}

@inproceedings{Swizzling,
author = {Kemper, Alfons and Kossmann, Donald},
title = {Adaptable Pointer Swizzling Strategies in Object Bases},
year = {1993},
isbn = {0818635703},
xpublisher = {IEEE Computer Society},
xaddress = {USA},
booktitle = {Proceedings of the Ninth International Conference on Data Engineering},
pages = {155–162},
numpages = {8}
}

@INPROCEEDINGS{LeanStore,
  author={Leis, Viktor and Haubenschild, Michael and Kemper, Alfons and Neumann, Thomas},
  booktitle={2018 IEEE 34th International Conference on Data Engineering}, 
  title={LeanStore: In-Memory Data Management beyond Main Memory}, 
  year={2018},
  volume={},
  number={},
  pages={185-196},
  xdoi={10.1109/ICDE.2018.00026}}

@electronic{im1,
  title         = "{Redis} - Real-time data for agents \& apps",
  year = {2026},
  url           = "https://redis.io/"
 }

@electronic{im2,
  title         = "{Volt Active Data} | Real-Time Decisioning Platform",
  year = {2026},
  url           = "https://www.voltactivedata.com/"
 }

@article{dbx1000-1,
  title={Staring into the abyss: An evaluation of concurrency control with one thousand cores},
  author={ Yu, Xiangyao  and  Bezerra, George  and  Pavlo, Andrew  and  Devadas, Srinivas  and  Stonebraker, Michael },
  journal={Proceedings of the VLDB Endowment},
  volume={8},
  number={3},
  year={2014},
  issue_date = {Nov 2014},
  xpublisher = {Association for Computing Machinery (ACM)},
  issn = {21508097},
  xurl = {http://hdl.handle.net/1721.1/100022},
  xdoi = {10.14778/2735508.2735511},
  month = nov,
  pages = {209–220},
  numpages = {12}
}

@inproceedings{dbx1000-2,
author = {Yu, Xiangyao and Pavlo, Andrew and Sanchez, Daniel and Devadas, Srinivas},
title = {TicToc: Time Traveling Optimistic Concurrency Control},
year = {2016},
isbn = {9781450335317},
xpublisher = {Association for Computing Machinery},
xaddress = {New York, NY, USA},
xurl = {https://doi.org/10.1145/2882903.2882935},
xdoi = {10.1145/2882903.2882935},
booktitle = {Proceedings of the 2016 International Conference on Management of Data},
pages = {1629–1642},
numpages = {14},
location = {San Francisco, California, USA},
xseries = {SIGMOD '16}
}

@article{dbx1000-3,
  title={Taurus: Lightweight Parallel Logging for In-Memory Database Management Systems (Extended Version)},
  author={ Xia, Yu  and  Yu, Xiangyao  and  Pavlo, Andrew  and  Devadas, Srinivas },
  journal={Proceedings of the VLDB Endowment},
  volume={14},
  number={2},
  year={2020},
  issue_date = {2020},
  xpublisher = {VLDB Endowment},
  xurl = {https://hdl.handle.net/1721.1/143468},
}

@article{dbx1000-4,
author = {Yu, Xiangyao and Xia, Yu and Pavlo, Andrew and Sanchez, Daniel and Rudolph, Larry and Devadas, Srinivas},
title = {Sundial: harmonizing concurrency control and caching in a distributed OLTP database management system},
year = {2018},
issue_date = {June 2018},
xpublisher = {VLDB Endowment},
volume = {11},
number = {10},
issn = {2150-8097},
xurl = {https://doi.org/10.14778/3231751.3231763},
xdoi = {10.14778/3231751.3231763},
journal = {Proceedings of the VLDB Endowment},
month = jun,
pages = {1289–1302},
numpages = {14}
}

@article{switch1,
author = {Al Maruf, Hasan and Chowdhury, Mosharaf},
title = {Memory Disaggregation: Advances and Open Challenges},
year = {2023},
issue_date = {June 2023},
xpublisher = {Association for Computing Machinery},
xaddress = {New York, NY, USA},
volume = {57},
number = {1},
issn = {0163-5980},
xurl = {https://doi.org/10.1145/3606557.3606562},
xdoi = {10.1145/3606557.3606562},
journal = {ACM SIGOPS Operating Systems Review},
month = jun,
pages = {29–37},
numpages = {9}
}

@inproceedings{switch2,
author = {Liu, Jinshu and Hadian, Hamid and Wang, Yuyue and Berger, Daniel S. and Nguyen, Marie and Jian, Xun and Noh, Sam H. and Li, Huaicheng},
title = {Systematic CXL Memory Characterization and Performance Analysis at Scale},
year = {2025},
isbn = {9798400710797},
xpublisher = {Association for Computing Machinery},
xaddress = {New York, NY, USA},
xurl = {https://doi.org/10.1145/3676641.3715987},
xdoi = {10.1145/3676641.3715987},
booktitle = {Proceedings of the 30th ACM International Conference on Architectural Support for Programming Languages and Operating Systems, Volume 2},
pages = {1203–1217},
numpages = {15},
location = {Rotterdam, Netherlands},
xseries = {ASPLOS '25}
}

@inproceedings {switch3,
author = {Yuhong Zhong and Fiodar Kazhamiaka and Pantea Zardoshti and Shuwei Teng and Rodrigo Fonseca and Mark D. Hill and Daniel S. Berger},
title = {Octopus: Enhancing {CXL} Memory Pods via Sparse Topology},
booktitle = {23rd USENIX Symposium on Networked Systems Design and Implementation},
year = {2026},
isbn = {978-1-939133-54-0},
xaddress = {Renton, WA},
pages = {1303--1322},
xurl = {https://www.usenix.org/conference/nsdi26/presentation/zhong},
xpublisher = {USENIX Association},
month = may
}

@article{nephele,
  author = {Cao, Huijie and Yang, Shun and Li, Yunfan and Huang, Chenfeng and Liu, Shengchi and Hu, Huiqi and Zhou, Xuan and Qian, Weining and Yang, Zhi and Yang, Hongshan},
  title = {{Nephele}: Bridging {DDR--CXL} Latency Gaps with Benefit-Aware Caching for In-Memory Databases on Hybrid Memory},
  journal = {ACM Transactions on Architecture and Code Optimization},
  year = {forthcoming},
  note = {Accepted for publication}
}

\end{document}